\documentclass[]{aa} 
\usepackage{graphicx} 
\usepackage{natbib}
\usepackage[utf8]{inputenc} 
\usepackage{txfonts} 
\usepackage{xcolor}

\newcommand{\emm}[1]{\ensuremath{#1}} 
\newcommand{\emr}[1]{\emm{\mathrm{#1}}}
\newcommand{\unit}[1]{\emr{\,#1}} 
\newcommand{\yr}{\unit{yr}}
\newcommand{\pc}{\unit{pc}} 

\newcommand{\cm}{\unit{cm}} 
\newcommand{\au}{\unit{a.u.}}
\newcommand{\mum}{\unit{\mu m}} 
\newcommand{\gpccm}{\unit{g\,cm^{-3}}}
\newcommand{\Kpccm}{\unit{K\,cm^{-3}}} 
\newcommand{\pccm}{\unit{cm^{-3}}}
\newcommand{\pscm}{\unit{cm^{-2}}}
\newcommand{\pscmpKkms}{\unit{cm^{-2}/(K\,km\,s^{-1})}}
 
\newcommand{\kms}{\unit{km\,s^{-1}}}
\newcommand{\Kkms}{\unit{K\,km\,s^{-1}}} 
\newcommand{\K}{\unit{K}}
 
\newcommand{\mm}{\unit{mm}}
\newcommand{\kHz}{\unit{kHz}} 
\newcommand{\MHz}{\unit{MHz}}
\newcommand{\GHz}{\unit{GHz}}

\newcommand{\Lsun}{\unit{L_{\sun}}} 
\newcommand{\Msun}{\unit{M_{\sun}}}
\newcommand{\Msunpyr}{\unit{M_{\sun} yr^{-1}}} 
\newcommand{\Av}{\emm{A_{v}}}
\newcommand{\Ebv}{\emm{E_\emr{B-V}}} 
\newcommand{\Xco}{\emm{X_\emr{CO}}}
\newcommand{\Wco}{\emm{W_\emr{CO}}} 
\newcommand{\Lco}{\emm{L_\emr{CO}}}
\newcommand{\magn}{\unit{mag}}

\newcommand{\chem}[1]{\ensuremath{\mathrm{#1}}} 
\newcommand{\Ht}{\chem{H_{2}}}
\newcommand{\Halpha}{\chem{H_{\alpha{}}}} 
\newcommand{\twCO}{\chem{^{12}CO}}
\newcommand{\thCO}{\chem{^{13}CO}} 
\newcommand{\Jtwo}{(2--1)}
\newcommand{\Jone}{(1--0)} 

\newcommand{\CHp}{\chem{CH^{+}}}

\newcommand{\Tsys}{\emm{T_\emr{sys}}} 
\newcommand{\Tas}{\emr{T_{A}^{*}}}
\newcommand{\Tmb}{\emr{T_{mb}}} 
\newcommand{\Feff}{\emr{F_{eff}}}
\newcommand{\Beff}{\emr{B_{eff}}}

\newcommand{\Mstar}{\emm{M_\star}} 
\newcommand{\Lstar}{\emm{L_\star}}
\newcommand{\Sstar}{\emm{S_\star}}

\newcommand{\Vlsr}{\emm{V_\emr{LSR}}} 
\newcommand{\Vstar}{\emm{V_\star}}
\newcommand{\Vwind}{\emm{V_\emr{wind}}} 
\newcommand{\Vgas}{\emm{V_\emr{g}}}
\newcommand{\Vdust}{\emm{V_\emr{d}}} 
\newcommand{\dV}{\emm{\Delta V_\emr{CO}}}

\newcommand{\Fdrag}{\vec{F_\emr{d}}} 
\newcommand{\Mrate}{\emm{\dot{M}}}

\newcommand{\Pther}{\emm{P_\emr{therm}}} 
\newcommand{\Pwind}{\emm{P_\emr{wind}}}
\newcommand{\Prad}{\emm{P_\emr{rad}}}

\newcommand{\Rstrom}{\emm{R_\emr{St}}} 
\newcommand{\Rcg}{\emm{R_\emr{CG}}}

\newcommand{\Robs}{\emm{R_\emr{obs}}} 
\newcommand{\Rapex}{\emm{R_\emr{apex}}}
\newcommand{\Rwind}{\emm{R_\emr{wind}}} 
\newcommand{\Rrad}{\emm{R_\emr{av}}} 

\newcommand{\Rthin}{\emm{R_\emr{rad}}}
\newcommand{\Rthick}{\emm{R^\prime_\emr{rad}}}

\newcommand{\namb}{\emm{n_\emr{a}}}
\newcommand{\nclump}{\emm{n_\emr{globulette}}}
\newcommand{\timecg}{\emm{t_\emr{CG}}}

\newcommand{\rhoamb}{\emm{\rho_\emr{a}}}
\newcommand{\rhowind}{\emm{\rho_\emr{w}}}
\newcommand{\rhogas}{\emm{\rho_\emr{g}}}
\newcommand{\rhodust}{\emm{\rho_\emr{d}}}
\newcommand{\rhobulk}{\emm{\rho_\emr{bulk}}}

\newcommand{\incli}{\emm{i}}

\newcommand{\sciexp}[2]{\emm{#1\times10^{#2}}}
\newcommand{\radec}[6]{\chem{\alpha_{2000}=#1^{h}#2^{m}#3^{s}},
\chem{\delta_{2000}=#4^{\circ}#5^{'}#6^{''}}} 
\newcommand{\N}[1]{\emm{N(#1)}}

\newcommand{\ie} {{\em i.e.}} 
\newcommand{\eg} {{\em e.g.}} 
\newcommand{\cf}
{{cf.}}

\newcommand{\nds}[1]{\emm{\displaystyle#1}} 

\newcommand{\paren}[1]{\nds{\left( #1 \right) }}

\newcommand{\bracket}[1]{\nds{\left[ #1 \right] }} 

\newcommand{\referee}[1]{{ #1}}

\newcommand{\FigContext}{ 
\begin{figure}
	\centering{}
	\includegraphics[width=\hsize{}]{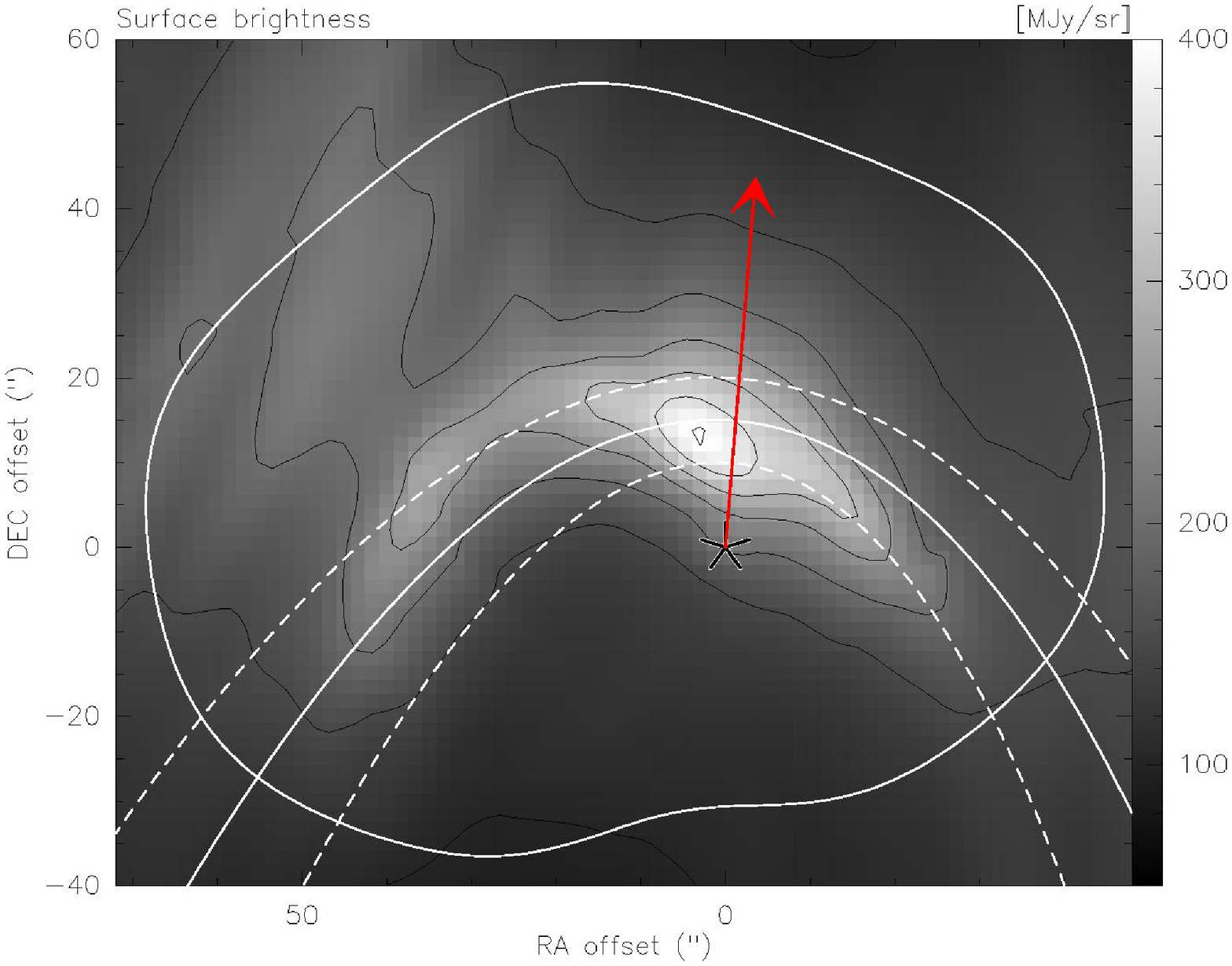} 
	\caption{Spitzer 24\mum{} image
	towards HD\,34078, from \citep{France.2007}. Contours are equally spaced every
	50\unit{MJy/sr}, starting from 100\unit{MJy/sr}. Offsets are measured from the
	star position, displayed as a black star. The red arrow represents the star
	on-sky proper motion for the next 1000 years. The solid open white curve
	displays a parabola pointing in this direction, with its focus at the star and a
	projected star-apex distance $\Robs = 15''$ (see Appendix~\ref{sec:projection}
	for a description of projection effects in a paraboloid). The two dashed curves
	correspond to variations of \Robs\ by $\pm 5\arcsec$. The closed white contour
	corresponds to the region mapped with the PdBI interferometer. {The right
	ascension axis increases towards the left.}} 
	\label{fig.context} 
\end{figure}
}

\newcommand{\FigICO}{ 
\begin{figure*}
	\centering{}
	\includegraphics[width=\hsize{}]{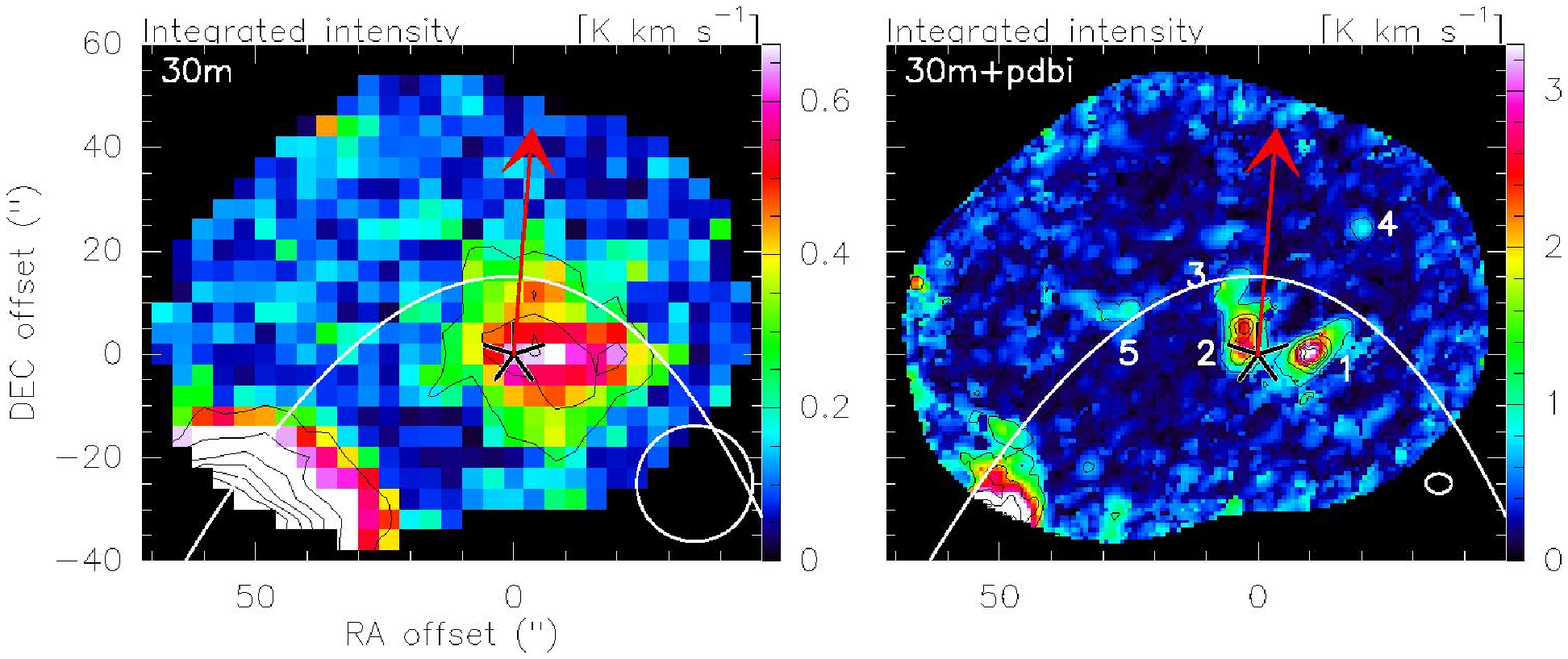} 
	\caption{Spatial distributions of
	the \twCO\Jone{} line integrated intensities. The left/right panels show the
	data from IRAM-30m only (at an angular resolution of $22.5''$), and the hybrid
	synthesis data from PdBI + IRAM-30m (at an angular resolution of $5.1''\times
	3.8''$ at a position angle of $139\degr$), respectively. The beam shape is drawn
	as an ellipse in the lower right corner of each panel. Black contours are drawn
	at 5, 10, 15, 20, 25 and 30 $\sigma$. Offsets are measured from the star
	position, denoted by the black star with the red arrow representing the star
	on-sky motion for the next 1000 years. The parabola adjusted to the 24\mum{} arc
	in Fig.~\ref{fig.context} with $\Robs = 15\arcsec$ is reproduced here to guide
	the eye. {The numbers 1-5 refer to the globulettes discussed in the
	text.}} 
	\label{fig.ICO} 
\end{figure*}
}

\newcommand{\FigMomCO}{ 
\begin{figure}
	\centering{}
	\includegraphics[height=23cm]{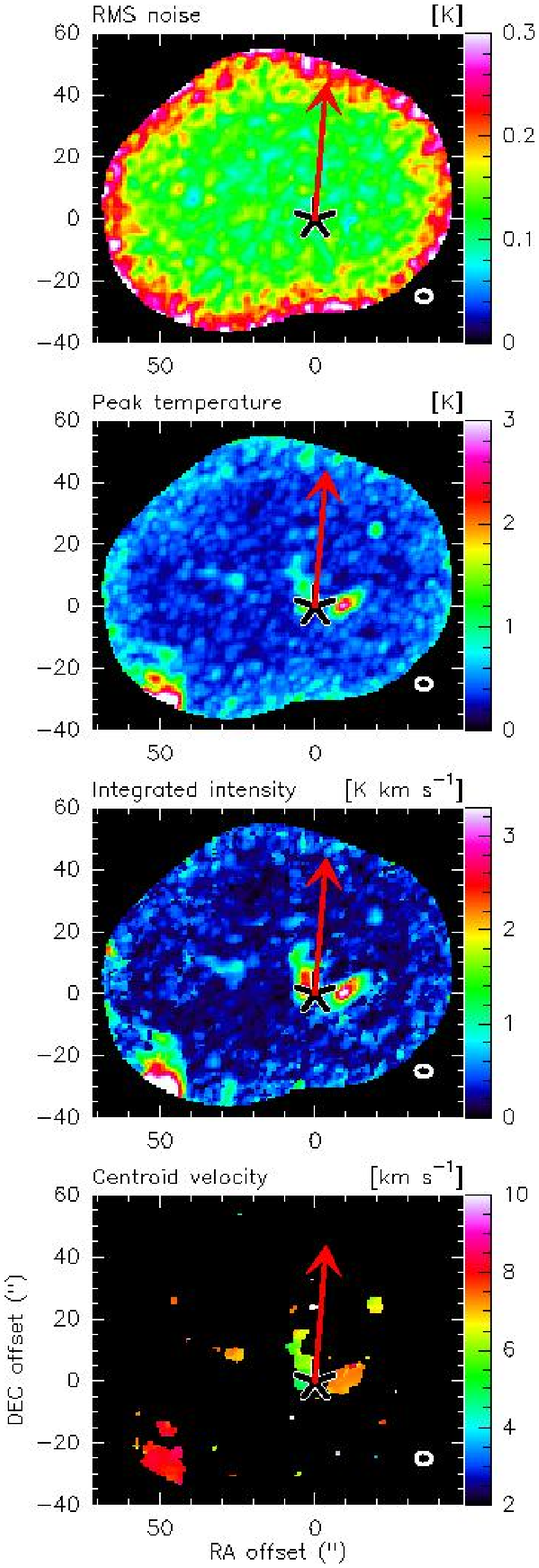} 
	\caption{Maps for the \twCO\Jone
	PdBI+30m data. From top to bottom RMS noise, peak temperature, integrated
	intensity and centroid velocity. In the last case, only the pixels with a peak
	SNR larger than 5 are shown.} 
	\label{fig.COmoms} 
\end{figure}
}

\newcommand{\FigSpectra}{ 
\begin{figure}
	\flushright{}
	\includegraphics[width=0.96\hsize{}]{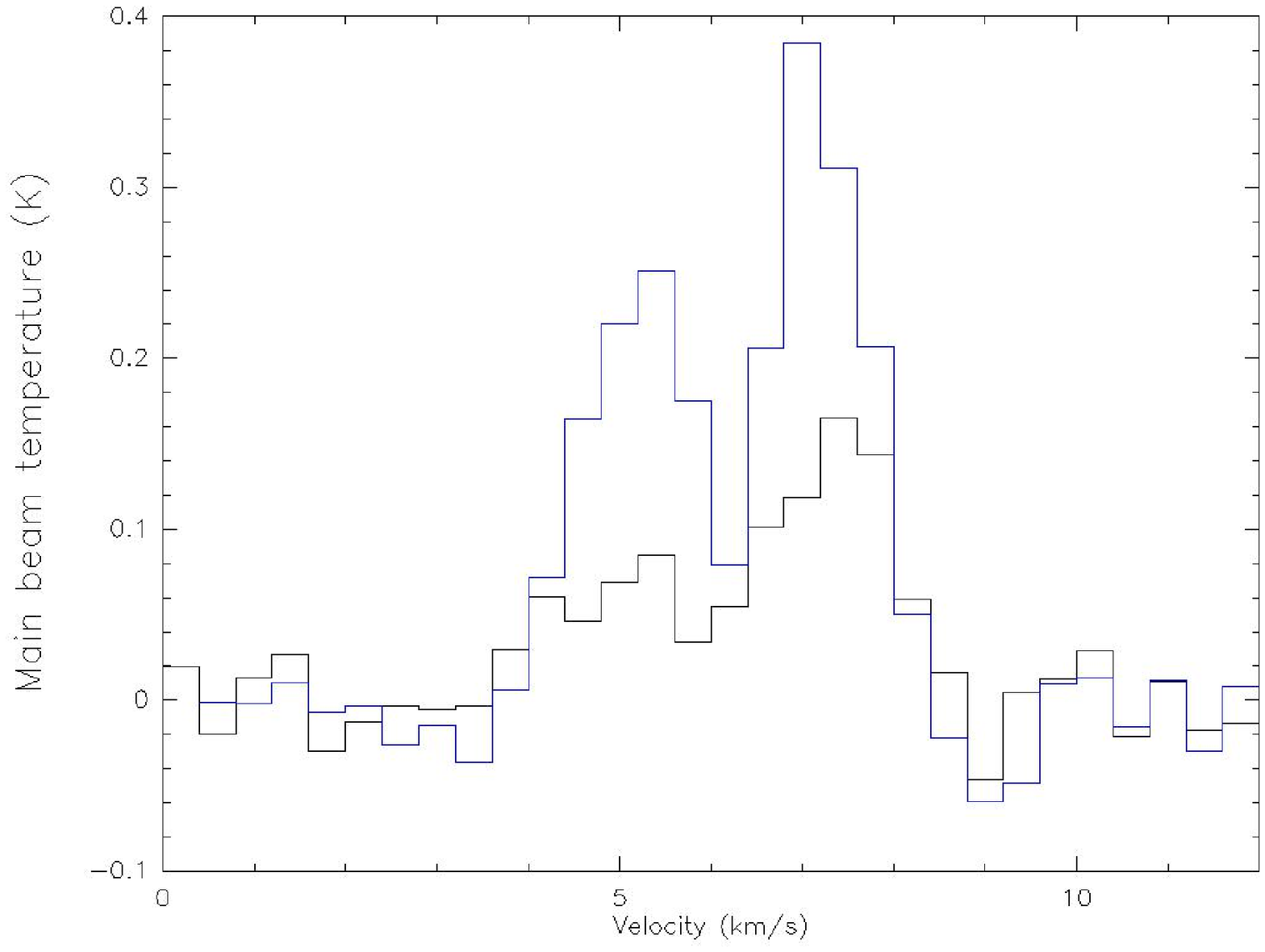}
	\includegraphics[width=\hsize{}]{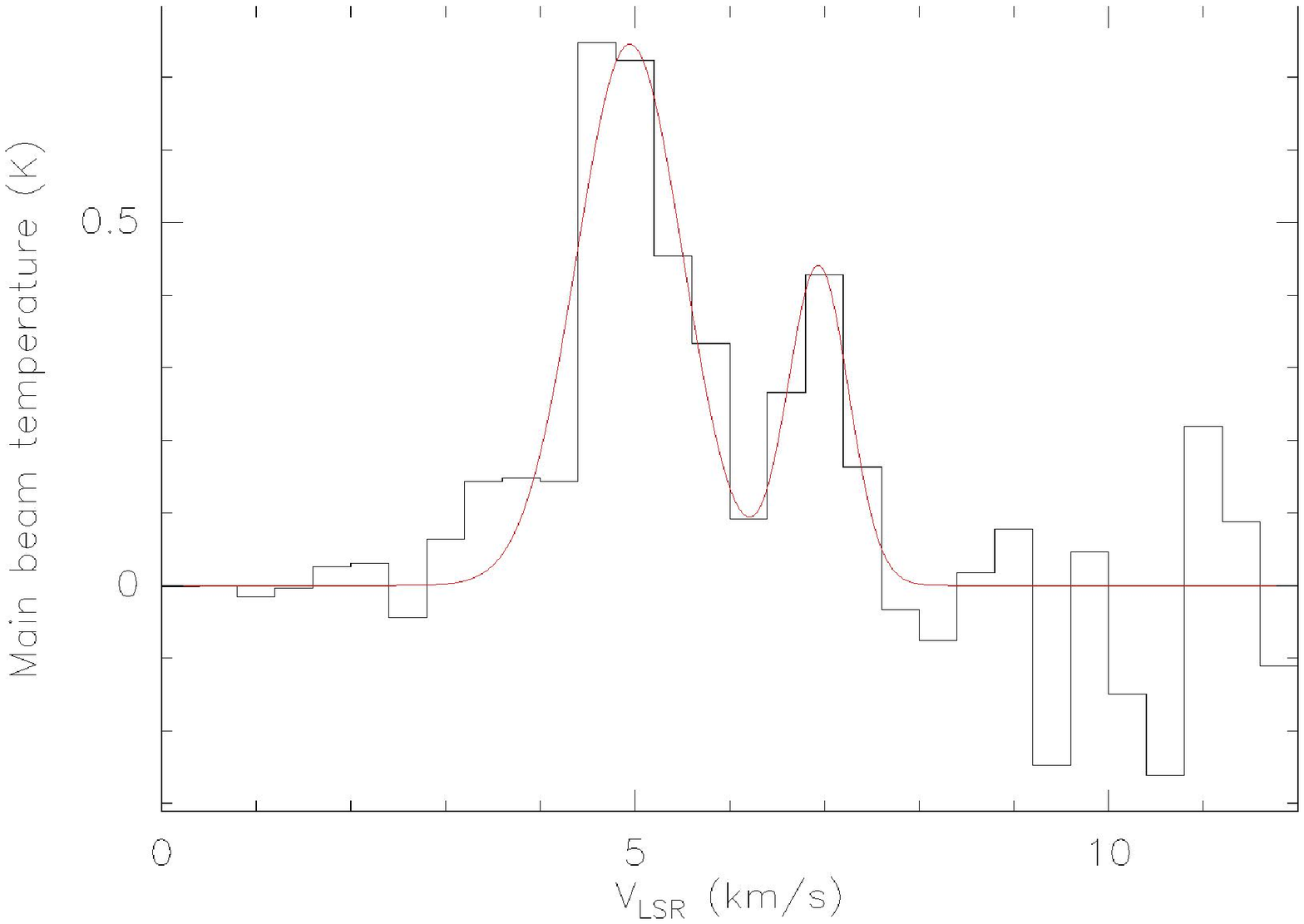} 
	\caption{Spectra observed
	towards the HD\,34078 star.
	\emph{Top:} \twCO\Jone{} (black) and \twCO\Jtwo{} (blue) spectra observed at
	the IRAM-30m telescope. The spectra are shown at their native resolution:
	$21.45''$ for \twCO\Jone\ and $10.7''$ for \twCO\Jtwo{}. \emph{Bottom:} Hybrid
	synthesis \twCO\Jone\ spectrum obtained at an angular resolution of $\sim4.4''$.
	The best fit, made with two gaussian components, is shown in red. The derived
	quantities are shown in Table~\ref{tab.LOS}.} 
	\label{fig.spectra} 
\end{figure}
}

\newcommand{\FigPV}{ 
\begin{figure}
	\centering{}
	\includegraphics[width=\hsize{}]{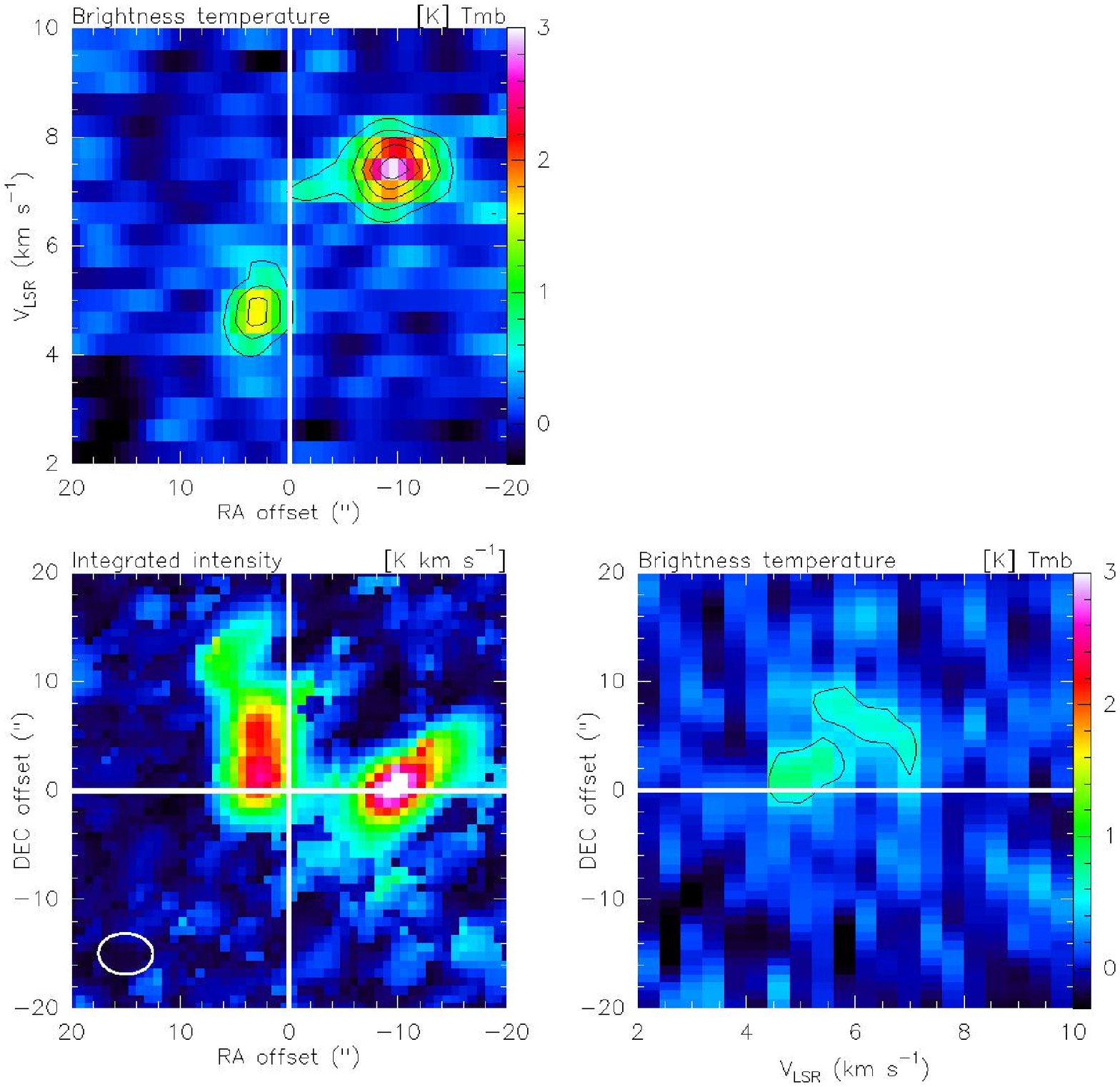} 
	\caption{Position velocity slice
	along the right ascention (upper left) and declination (lower right) axis, going
	trough the line of sight to HD\,34078. The vertical and horizontal white lines
	corresponds to the star position. {The bottom left panel is the
	integrated CO intensity.}} 
	\label{fig.pv} 
\end{figure}
}

\newcommand{\FigProjection}{ 
\begin{figure}
	\centering{}
	\includegraphics[width=\hsize{}]{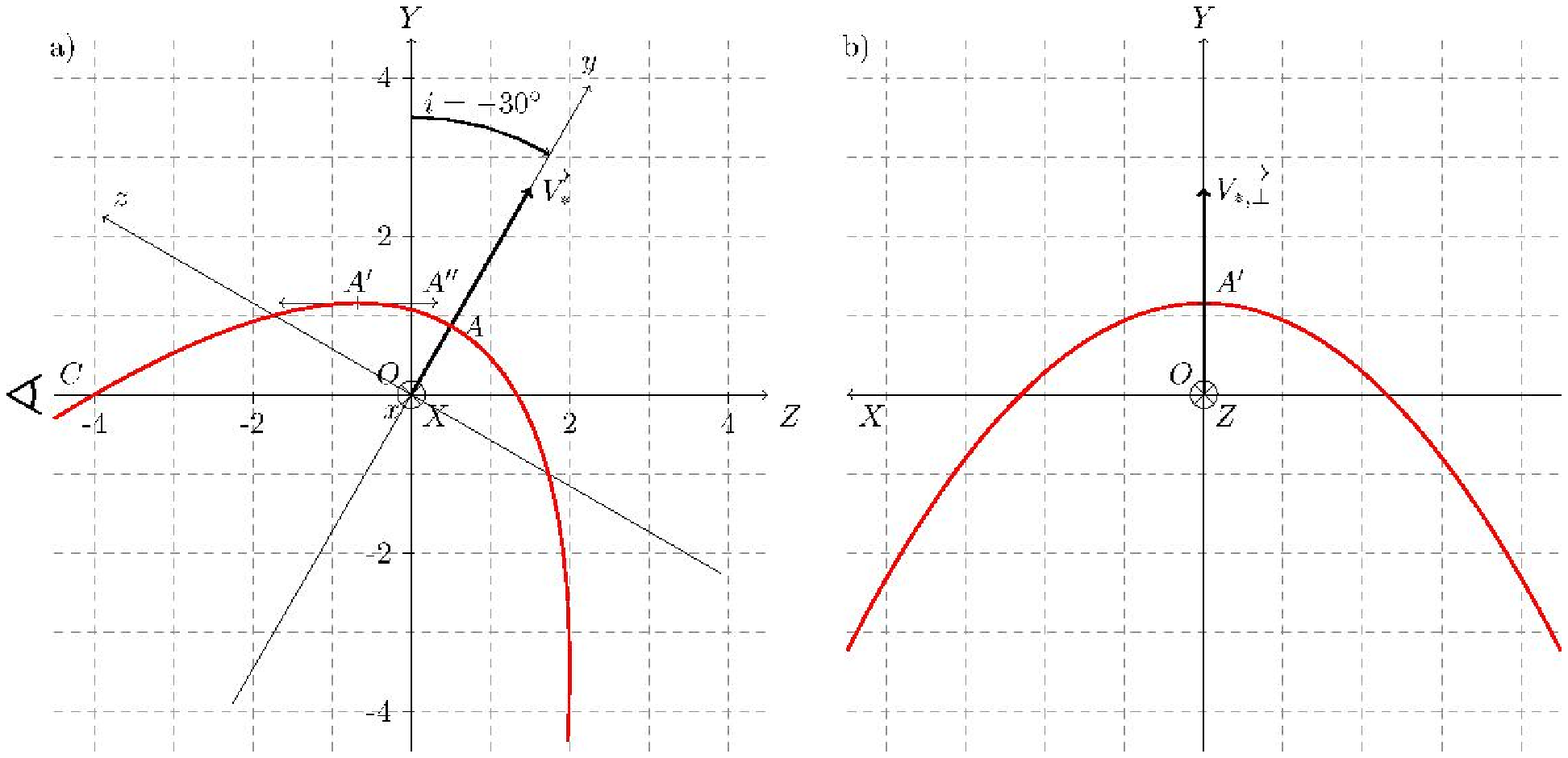} 
	\caption{Projections of
	a paraboloid shape from the natural star frame to the observer frame. Panel (a)
	shows a cut through the paraboloid in the plane $x=X=0$ defined by the star
	proper motion vector $\vec{V_*}$ and the line of sight (Z axis pointing away
	from the observer). Point A is the real apex of the parabola, A$'$ is the
	apparent apex as seen by the observer, A$''$ the projection of A$'$ on the Y
	axis, and C the intersection of the front side of the parabola with the line of
	sight. Panel (b) shows the paraboloid apparent shape as seen in projection in
	the XY plane of the sky.} 
	\label{fig.projection} 
\end{figure}
}

\newcommand{\TabObservations}{ 
\begin{table*}
	\caption{
	\label{tab.obs}Observation parameters. The projection center for the all the
	observations is \radec{05}{16}{19.143}{34}{18}{52.34} the position of
	HD\,34078.} 
	\begin{center}
		\small 
		\begin{tabular}
			{rccccccccccc} 
			\hline Molecule
			& Transition & Frequency & Instr. & Config. & Beam & PA & Vel. res. & Int.
			Time 
			\tablefootmark{a} & Tsys & Noise 
			\tablefootmark{b} & Obs. date \\
			& &
			\GHz) & & & ($''$) & ($\deg$) & (\kms) & (hr) & (K) & (mK) &\\
			\hline \twCO &
			$J=$ \Jone &$115.27120$& PdBI & 5D & $5.1\times3.8$ &$ 139 $&$ 0.2 $&$ 8.5/19
			$&$ 200 $ & 140 & Aug. 2011\\
			\hline 
		\end{tabular}
		\begin{tabular}
			{rcccccccccccc} 
			\hline Molecule & Transition & Frequency & Instr. & \#pol &
			\Feff{} & \Beff{} & Res. & Res. & Int. Time & Tsys & Noise 
			\tablefootmark{b} &
			Obs. date \\
			& & (\GHz) & & & & & ($''$) & (\kms) & (hr) & (K) & (mK)&\\
			\hline
			\twCO & $J=$ \Jtwo &$230.53800$& 30m/E2 & 2 & $0.91$ &$ 0.59 $&$ 11.2 $ & $0.4
			$&$ 7.1/20 $&$ 460 $ & 140 & May 2012\\
			\twCO & $J=$ \Jone &$115.27120$& 30m/E0
			& 2 & $0.95$ &$ 0.78 $&$ 22.5$ & $0.4 $&$ 7.1/20 $&$ 230 $ & 65 & May 2012\\

			\hline 
		\end{tabular}
		\tablefoot{ 
		\tablefoottext{a}{listed as on-source
		time/telescope time} 
		\tablefoottext{b}{evaluated at the mosaic phase center (the
		noise steeply increases at the mosaic edges after correction for primary beam
		attenuation, see top panel of Fig. \ref{fig.COmoms})}} 
	\end{center}
\end{table*}
}

\newcommand{\TabClumpProps}{ 
\begin{table*}
	\caption{
	\label{tab.clumpprops}Observed and inferred properties of the CO globulettes, as
	numbered in the right panel of Fig.~\ref{fig.ICO}.} 
	\begin{center}
		\tiny
		\begin{tabular}
			{lcccccccccccc} 
			\hline \# & Position & \unit{V_{LSR}} & Size &
			\dV & $<\Wco>$ & $\Wco^\emr{peak}$ & \Av 
			\tablefootmark{a,b} & \Lco{}
			\tablefootmark{c} & Mass from CO 
			\tablefootmark{a,e} & Virial Mass & \nclump
			\tablefootmark{f} \\
			& ($''$) & (\kms) & ($''$) & (\kms) & (\Kkms) & (\Kkms) &
			(\magn) & (\Kkms\,\pc$^2$) & (\Msun) & (\Msun) & (\pccm)\\
			\hline 1 & (-9.9,0.3)
			& 7.4 & $9.2\times4.1$ & 1.3 & 1.03 & 3.70 & 0.82 & $1.39\times10^{-3}$ &
			$6.21\times10^{-3}$ & $5.0$ & $9\times10^{4}$ \\
			2 & (3.8, 3.3) & 5.5 &
			$10.6\times5.1$ & 2.6 & 1.18 & 2.82 & 0.63 & $1.27\times10^{-3}$ &
			$5.71\times10^{-3}$ & $24.1$ & $5\times10^{4}$ \\
			3 & (4.5, 13.3) & 6.1 &
			$7.0\times4.1$ & 2.6 
			\tablefootmark{d} & 0.78 & 1.76 & 0.39 &
			$3.30\times10^{-4}$ & $1.48\times10^{-3}$ & $17.6$ & $3\times10^{4}$ \\
			4 &
			(-20.2, 24.9) & 6.6 & $5.9\times4.6$ & 0.6 & 0.52 & 0.85 & 0.19 &
			$1.18\times10^{-4}$ & $5.28\times10^{-4}$ & $0.9$ & $1\times10^{4}$ \\
			5 &
			(29.45, 8.5) & 7.3 & $15.2\times6.8$ & 1.1 & 0.46 & 1.06 & 0.24 &
			$5.04\times10^{-4}$ & $2.26\times10^{-3}$ & $6.0$ & $1\times10^{4}$ \\
			\hline
		\end{tabular}
		\tablefoot{ 
		\tablefoottext{a}{Computed with a standard \Xco{}
		conversion factor $= 2\times10^{20}\pscmpKkms$ and $N_\emr{H}/\Av =
		1.8\times10^{21}\pscm/\magn$.} 
		\tablefoottext{b}{This \Av{} is computed from
		$\Wco^\emr{peak}$.} 
		\tablefoottext{c}{Assuming a distance to HD\,34078 of
		530\pc{}, $1''$ corresponds to $2.6\times10^{-3}\pc$.} 
		\tablefoottext{d}{The
		spectrum towards this globulette is double peaked, the line width estimate is an
		upper limit.} 
		\tablefoottext{e}{Including Helium.} 
		\tablefoottext{f}{Hydrogen
		density computed assuming a spherical cloud with a diameter equal to the
		geometrical mean of the major and minor axes given in column 4 of this table.} }
	\end{center}
\end{table*}
}

\newcommand{\TabLOSProps}{ 
\begin{table}
	\begin{center}
		\caption{
		\label{tab.LOS}Properties of the two spectral components in the direction of
		HD\,34078 derived from PdBI+30m data with $4.4''$ angular resolution.}
		\begin{tabular}
			{lcc} 
			\hline Spectral component & 1 & 2\\
			\hline \Vlsr{} (\kms)
			& $4.9 \pm 0.1$ & $6.9 \pm 0.2$ \\
			\dV{} (\kms) & $1.3 \pm 0.2$ & $0.8 \pm 0.7$
			\\
			\Wco{} (\Kkms) & $1.1 \pm 0.3$ & $0.4 \pm 0.4$ \\
			\Av{} (mag) & $0.23 \pm
			0.06$ & $0.08 \pm 0.08$ \\
			\hline 
		\end{tabular}
	\end{center}
\end{table}
}

\newcommand{\TabParameters}{ 
\begin{table*}
	\caption{
	\label{tab.parameters}Parameter symbols, nominal values, and scalings in
	HD\,34078 and its bow shock} 
	\begin{center}
		\resizebox{19cm}{!}{ 
		\begin{tabular}
			{lccllc} 
			\hline Definition & Symbol & Eq. & Nominal value & Scaling &
			Reference\\
			\hline Distance & $D$ & & $530 \pc$ & & (1)\\
			Spectral type & --- &
			& O9.5 & & (2)\\
			Stellar mass & \Mstar{} & & $20\Msun$ & & (2)\\
			Stellar
			luminosity & \Lstar{} & & ${8}\times 10^{4}\Lsun$ & $\propto D^2$ & ({1})\\
			Stellar velocity & \Vstar{} & & $150_{-50}^{+30}\kms$ & $\propto D$ & (3)\\
			Wind
			velocity & \Vwind{} & & $800\kms$ & & (2)\\
			Wind mass loss rate (from UV) &
			\Mrate{} & & $3 \times 10^{-10}\Msunpyr$ & & (2)\\
			\hline Str{\"o}mgren radius &
			\Rstrom{} & (\ref{eq.stromgren}) & $4.3\pc$ & $\propto D$ & (4)\\
			Ambient
			density & \namb{} & (\ref{eq.stromgren}) & ${ 20}\pccm$ & $\propto D^{3/2}$ &
			(4)\\
			Wind ram pressure & \Pwind{} & (\ref{eq.pwind}) & ${ 2\,300}\Kpccm$ at
			$R=0.2\pc$ & $\propto \Mrate\Vwind R^{-2}$ & (4)\\
			Radiation pressure & \Prad{}
			& (\ref{eq.prad}) & ${ 1.5}\times 10^{7}\Kpccm$ at $R=0.2\pc$ & $\propto D^2
			R^{-2}$ & (4)\\
			Globulette thermal pressure & \Pther{} & (\ref{eq.ptherm}) &
			$\sim 10^6-10^7\Kpccm$ & $\propto D^{-1} T$ & (4)\\
			\hline Star-apex distance
			(Observed) & \Robs{} & & $0.04\pc$ & $\propto D$ & (4)\\
			Star-apex distance
			(Deprojected) & \Rapex{} & & $0.035\pc$ & $\propto D$ & (4)\\
			Wind standoff
			radius & \Rwind{} & (\ref{eq.Rwind}) & ${ 0.0012}\pc$ & $\propto
			(\Mrate\Vwind)^{1/2} D^{-7/4}$ & (4)\\
			Dust avoidance radius ( no gas drag) &
			\Rrad{} & (\ref{eq.avoidance}) & ${ 0.05-0.7}\pc$ & $\propto D^{0}$ & (4)\\
			Gas+grain avoidance radius (opt. thin) & \Rthin{} & (\ref{eq.rthin}) &
			$\overline{\Rrad}/100 = { 0.0024} \pc$ & $\propto D^{0}$ & (4)\\
			Gas+grain
			avoidance radius (opt. thick) & \Rthick{} & (\ref{eq.rthick}) & ${ 0.10}\pc$ &
			$\propto D^{-3/4}$ & (4)\\
			\hline 
		\end{tabular}
		} 
		\tablefoot{ References: (1)
		\citet{Herbig.1999}, (2) \citet{Martins.2005}, (3) \citet{Tetzlaff.2011}, (4)
		This work.} 
	\end{center}
\end{table*}
}

\begin{document} 

\title{Dense molecular ``globulettes'' and the dust arc\\
towards the runaway O
star AE\,Aur (HD\,34078)\thanks{Based on observations carried out with the IRAM
Plateau de Bure Interferometer. IRAM is supported by INSU/CNRS (France), MPG
(Germany) and IGN (Spain).}}

\author{P. Gratier \inst{\ref{IRAM},\ref{LAB}, \ref{CNRS}} \and J. Pety
\inst{\ref{IRAM},\ref{LERMA}} \and P. Boiss\'e \inst{\ref{IAP}} \and S.
Cabrit \inst{\ref{LERMA}} \and P. Lesaffre \inst{\ref{LERMA},\ref{Vietnam}} \and
M. Gerin \inst{\ref{LERMA}} \and G.
Pineau des For\^ets\inst{\ref{IAS},\ref{LERMA}}}

\institute{ IRAM, 300 rue de la Piscine, 38406 Saint Martin d'H\`eres, France \\
\email{[gratier,pety]@iram.fr} 
\label{IRAM} \and Univ. Bordeaux, LAB, UMR 5804,
F-33270, Floirac, France 
\label{LAB} \and CNRS, LAB, UMR 5804, F-33270, Floirac,
France 
\label{CNRS} \and LERMA, UMR 8112, CNRS, Observatoire de Paris, ENS,
UPMC, UCP, 61 avenue de l'Observatoire, F-75014 Paris 
\label{LERMA} \and IAP,
UMR7095 CNRS and Universit\'e Pierre et Marie Curie - Paris 6, 98bis boulevard
Arago, 75014, Paris, France 
\label{IAP} \and VATLY, Institute for Nuclear
Science and Technology, 179 Hoang Quoc Viet, Cau Giay, Ha Noi, Viet Nam
\label{Vietnam} \and IAS, UMR 8617, CNRS, B{\^a}timent 121, Universit\'e Paris
Sud 11, 91405, Orsay, France 
\label{IAS}}

\date{}

\abstract {Some runaway stars are known to display IR arc-like structures around
them, resulting from their interaction with surrounding interstellar material.
The properties of these features as well as the processes involved in their
formation are still poorly understood.} {\referee{We aim at understanding the
physical mechanisms that shapes the dust arc observed near the runaway O-star
AE\,Aur (HD\,34078).}} {\referee{We obtained and analyzed a high spatial
resolution ($4.4''$) map of the $^{12}$CO(1-0) emission that is centered on
HD\,34078, and that combines data from both the IRAM interferometer and 30m
single-dish antenna.}} {One third of the 30m flux mainly originates from two
small (no larger than $5''\times10''$ or $0.013 \times 0.026\pc$), and bright (1
and 3\K{} peak temperatures) CO globulettes. The line of sight towards HD\,34078
intersects the outer part of one of them, which accounts for both the properties
of diffuse UV light observed in the field by~\citet[]{France.2004} and the
numerous molecular absorption lines detected in HD\,34078's spectra, including
those from highly excited \Ht{}. Their modelled distance from the star
\referee{(0.2\pc)} is compatible with the fact that they lie on the 3D
paraboloid which fits the arc detected in the 24\mum{} Spitzer image. Four other
compact CO globulettes are detected in the mapped area, all lying close to the
rim of this paraboloid. These globulettes have a high density and linewidth, and
are strongly pressure-confined or transient.} {The presence of molecular
globulettes at such a close distance from an O star is unexpected, and probably
related to the high proper motion of HD\,34078. Indeed, the good spatial
correlation between the CO globulettes and the IR arc suggests that they result
from the interaction of the radiation and wind emitted by HD\,34078 with the
ambient gas. However, the details of this interaction remain unclear. A wind
mass loss rate significantly larger than the value inferred from UV lines is
favored by the large IR arc size, but does not easily explain the low velocity
of the CO globulettes. The effect of radiation pressure on dust grains also
meets several issues in explaining the observations.
Further observational and theoretical work is needed to fully elucidate the
processes shaping the gas and dust in bow shocks around runaway O stars.}

\maketitle{} 
\section{Introduction}

\TabObservations{} \FigContext{} The interstellar medium (ISM) was discovered in
1907 by studying atomic absorption lines (from NaI and CaII) seen in the visible
spectrum of bright stars. The first interstellar molecules (CN, CH, and \CHp{})
were also detected in the same way in the years 1937--1940. In the seventies,
the Copernicus satellite systematically studied \Ht{} ultraviolet (UV)
absorption lines to investigate the properties of diffuse interstellar gas with
low visual extinction ($\Av \le 1\magn$), and this powerful method is still
largely used today (\cf{} FUSE, HST/STIS and HST/COS programs). The gas in
diffuse clouds is mainly neutral, warm (typically 80\K{}), and (usually) of low
density ($100-500\pccm $). Such regions correspond to the transition from atomic
to molecular hydrogen, where carbon is still mostly ionised or neutral, with
$\N{\emr{CO}} < \mbox{a few~} 10^{16}\pscm$, and $\N{\emr{C}} \sim
3.10^{17}\pscm$.

In this framework, the foreground absorption against the O9.5 star HD\,34078
stands out for its peculiar properties. HD\,34078 was ejected about 2.5\,Myr ago
from the Orion region \citep[]{Blaauw.1953,Bagnuolo.2001} and is now the fastest
runaway star in the local ISM, with a velocity of $\Vstar \sim 150\kms$
\citep{Tetzlaff.2011}. The line of sight towards HD\,34078 should thus offer a
means to detect small scale $(5-50\au)$ density or abundance variations in the
diffuse interstellar medium in only a few decades. However, as absorption line
studies progressed, it was realized that this line of sight exhibits peculiar
properties compared to the usual diffuse ISM on other lines of sight. Very
abundant CH and \CHp{} is measured, with some time variability
\citep{Rollinde.2003}. The direct starlight suffers larger reddening than
scattered light in the surrounding nebula \citep{France.2004}. In addition, UV
absorption studies with FUSE reveal an unusually large amount of highly excited
\Ht{}, indicating the unexpected presence of dense ($n_H \simeq 10^4\pccm$) and
strongly irradiated molecular gas at about 0.2\pc{} from the
star~\citep{Boisse.2005}. The latter property could be related to the recent
interaction of the star with the IC\,405 reflection nebula \citep{Herbig.1958}.
Indeed, the emission of hot dust at 24\mum{} imaged with Spitzer/MIPS clearly
delineates a parabolic curve (see Fig.~\ref{fig.context}), interpreted as the
tip of a bow shock resulting from the interaction of the fast stellar wind with
the preexisting diffuse gas of IC\,405 \citep{France.2007}.

In order to gain 2D information on the molecular gas structure and kinematics in
this bow shock, \citet{Boisse.2009} conducted sensitive \twCO{}\Jtwo{} mapping
in a narrow field of view around HD\,34078 using the IRAM-30m telescope at an
angular resolution of $12''$. On top of a widespread CO component, brighter
emission was detected in a clumpy ``filament'' peaking slightly below the
24\mum{} arc, confirming ongoing interaction between the star and the
surrounding cloud. However, the CO velocity field showed a gradient mainly
perpendicular to the star-apex axis, instead of mainly along this axis as
expected for a steady-state wind bow shock~\citep[see e.g.][]{Wilkin.1996}. In
addition, \citet{Boisse.2009} noted that the apparent distance between the star
and the IR arc of $15''$, corresponding to 0.04\pc{} at the distance of
HD\,34078~\citep[$\simeq$ 530\pc{} { assuming $M_V = -4.2$}][]{Herbig.1999}, is
largely incompatible with the prediction for a stationary bow shock for the wind
mass-flux derived from UV line analysis \citep{Martins.2005}. They then proposed
that we might be witnessing the recent birth of the bow shock, or that radiation
pressure on grains might play a role in increasing its size.
Alternatively, the wind mass-flux might have been largely underestimated (as
proposed by \citet{Gvaramadze.2012} to explain the IR bow shock size in
$\zeta$\,Oph) or some other under-appreciated physical process could be at play.

Distinguishing between these hypotheses is crucial in order to better understand
the processes that govern the interaction between HD\,34078 and the surrounding
ISM, how they impact the absorbing gas properties on the line of sight, and how
they may affect the wind mass-flux determinations from IR bow shock sizes in
other runaway O stars. As a step towards this goal, we mapped the bowshaped IR
arc around HD\,34078 in \twCO{}\Jone{} with the Plateau de Bure Interferometer
(PdBI, complemented with IRAM-30m single-dish data to provide the
short-spacings) at $\simeq 4''$ resolution, comparable to the 24\mum{} Spitzer
image and 3 times better than the \twCO{}\Jtwo{} map of~\citet{Boisse.2009}.

The observations and data reduction are presented in
Sect.~\ref{sec:observations}. The resulting properties of the detected CO
structures are described in Sect.~\ref{sec:results}. We discuss their
implications in Sect.~\ref{sec:discussion}. We summarize and conclude in Sect.~
\ref{sec:conclusions}.

\section{Observations and data reduction} 
\label{sec:observations}

\FigICO{} Table~\ref{tab.obs} summarizes the interferometric and single-dish
observations described in this section.

\subsection{Interferometric observations and data reduction}

Interferometric observations of HD\,34078 were obtained at the Plateau de Bure
Interferometer (PdBI) operated by IRAM. These observations where carried out
with 5 antennas in the D configuration (baselines from 24 to 96 meters) in
August 2011. We observed a mosaic of 13 pointings that followed an hexagonal
compact pattern with closest neighbors separated by half the primary beam. The
mosaic thus covers a field of view of $120''\times 100''$. We used the D
configuration of the array, yielding a typical synthesized angular resolution of
$4.4''$. The observations include about 19 hours of telescope time. The
on-source time scaled to a 6 antenna array is 8.5 hours. Three correlator
windows of 40\,MHz were concatenated to form a continuous bandwidth of 300\kms\
centered on the \twCO\Jone\ rest frequency (115.271\GHz) at a resolution of
0.2\kms. During the observations, the typical precipitable water vapor amounted
to 4-8mm and the typical system temperature was 200\K. The median noise level
achieved over the mosaic is 0.14\K{} (\Tmb{}) in channels of 0.4\kms{} width.

We used the standard algorithms implemented in the \texttt{GILDAS/CLIC} software
to calibrate the PdBI data. The radio-frequency bandpass was calibrated by
observing the bright (9 Jy) quasar 3C454.3. Phase and amplitude temporal
variations were calibrated by fitting spline polynomials through regular
mesurements of two nearby ($<12\degr$) quasars (J0418+380 and 0548+398). The
PdBI secondary flux calibrator MWC\,349 was observed once during every track,
which allowed us to derive the flux scale of the interferometric data. The
absolute flux accuracy is $\sim 10\%$.

\subsection{Single dish mapping observations and data reduction}

A multiplicative interferometer filters out the low spatial frequencies, \ie{},
spatially extended emission. We thus observed the same region with the IRAM-30m
single dish telescope in order to recover the low spatial frequency (``short-
and zero-spacing'') information filtered out by the PdBI. We describe here the
observing strategy and the calibration, baselining and gridding methods we used
to obtain single-dish data whose quality matches the interferometric data.

The single dish observations were taken at the IRAM-30m telescope in May 2012
during 20 hours of average summer time (9\mm{} median water vapor,
$\Tsys\sim230\K$ at 3\mm{} and $\Tsys\sim460\K$ at 1\mm{}). We observed
simultaneously at 3 and 1\,mm with a combination of the EMIR receivers and the
Fourier transform spectrometers, which yields a bandwidth of 3.6\GHz{} per
polarization at a frequency resolution of 49\kHz{}. This allowed us to measure
\twCO\Jone, \twCO\Jtwo, \thCO\Jone, and half of the CN\Jone{} hyperfine lines.

We used the on-the-fly scanning strategy with a dump time of 0.4 seconds and a
scanning speed of $5.5''/$s to ensure a sampling of 5 dumps per beam at the
$11''$ resolution of the \twCO\Jtwo\ line. The $300''\times250''$ map was
covered using successive orthogonal scans along the RA and DEC axes. The
separation between two successive rasters was $4.3''$ ($\sim \lambda/D$) to
ensure Nyquist sampling. A common off reference position located at offsets
($-400''$,$-200''$) was observed for 10 seconds every 40 to 50 seconds. The
calibration parameters (including the system temperature) were measured every 20
minutes using the hot/cold/sky load method. The pointing was checked every hour
and the focus every 4 hours.
The typical IRAM-30m position accuracy is $\sim 2''$.

Data reduction was carried out using the \texttt{GILDAS/CLASS} software. A
20\MHz{}-wide subset of the spectra was first extracted around each line's rest
frequency. We then computed the experimental noise by subtracting a zeroth order
baseline from every spectrum. A systematic comparison of this noise value with
the theoretical noise computed from the system temperature, the integration
time, and the channel width, allowed us to filter out outlier spectra. This
amounted to a few tens over the 212\,700 spectra of the full data set. The
spectra where then gridded to a data cube through a convolution with a Gaussian
kernel of FWHM$\sim1/3$ of the telescope beamwidth. Finally, we fitted another
baseline of order 3 through each spectrum of the cube. The two steps of
baselining excluded a velocity range of 0 to 10\kms{} LSR, where the signal
resides.

\subsection{Joint imaging and deconvolution of the interferometric and
single-dish data}

Following~\citet{Rodriguez-Fernandez.2008a}, the \texttt{GILDAS/MAPPING}
software and the single-dish map from the IRAM-30m were used to create the
short-spacing visibilities not sampled by the Plateau de Bure interferometer. In
short, the maps were deconvolved from the IRAM-30m beam in the Fourier plane
before multiplication by the PdBI primary beam in the image plane. After a last
Fourier transform, pseudo-visibilities were sampled between 0 and 15~m (the
diameter of the PdBI antenna). These visibilities were then merged with the
interferometric observations. Each mosaic field was imaged and a dirty mosaic
was built combining those fields in the following optimal way in terms of
signal--to--noise ratio~\citep{Pety.2010} :
\begin{displaymath}
	\displaystyle J(\alpha,\delta) = \sum\nolimits_i
	\frac{B_i(\alpha,\delta)}{\sigma_i^2}\,F_i(\alpha,\delta) \left/ \displaystyle
	\sum\nolimits_i \frac{B_i(\alpha,\delta)^2}{\sigma_i^2}.
	\right.
\end{displaymath}
In this equation, $J(\alpha,\delta)$ is the brightness
distribution in the dirty mosaic image, $B_i$ are the response functions of the
primary antenna beams, $F_i$ are the brightness distributions of the individual
dirty maps and $\sigma_i$ are the corresponding noise values. As can be seen in
this equation, the dirty intensity distribution is corrected for primary beam
attenuation, which induces a spatially inhomogeneous noise level. In particular,
noise strongly increases near the edges of the field of view.
To limit this effect, both the primary beams used in the above formula and the
resulting dirty mosaics are truncated. The standard level of truncation is set
at 20\% of the maximum in MAPPING. The dirty image was deconvolved using the
standard H\"ogbom CLEAN algorithm. The resulting data cube was then scaled from
Jy/beam to \Tmb{} temperature scale using the synthesized beam size (see
Table~\ref{tab.obs}).

\TabLOSProps{} \TabClumpProps{} 
\section{Observational results}
\label{sec:results}

\FigMomCO{} \FigPV{} \FigSpectra{} Figure~\ref{fig.ICO} compares the spatial
distributions of the \twCO\Jone\ integrated emission at the resolutions of the
IRAM-30m and PdBI instruments. The IRAM-30m map was masked to display exactly
the same field of view as the hybrid (PdBI + IRAM-30m) synthesis map. The noise
is non-uniform in both maps. A $50''$-square patch centered on the star position
was observed longer at the IRAM-30m to improve the signal-to-noise ratio in this
region. In addition, the noise naturally increases at the edges of the mosaic
because of the correction of the interferometer primary beam attenuation.

Two unresolved bright ``globules'' clearly pop up at the resolution of the
IRAM-30m. The brightest one, located to the south-east of the star, is only
partly included inside the observed field of view. We thus saturate the
corresponding region in the color scale of Fig.~\ref{fig.ICO} in order to
outline the structure of the second bright emitting region close to the star,
previously studied with the IRAM-30m by~\citet{Boisse.2009}. We find that, at
the higher resolution of the hybrid map, this ``globule'' clearly breaks up into
two bright (up to 3\Kkms{}) globulettes denoted as \#1 and \#2 in
Fig.~\ref{fig.ICO}, and peaking less than 10\arcsec{} on either side from the
star sightline. They are clearly elongated, with a typical length of $\sim10''$
and a width of $\sim 4''-5''$ (\ie{}, barely resolved) in the perpendicular
direction. In both cases, the elongation is roughly pointing in the direction of
the star.

Three other globulettes are detected in the hybrid synthesis map with a
signal-to-noise ratio larger than 5. These lower brightness $(\la 1\Kkms)$
structures, denoted as \#3, \#4 and \#5 in Fig.~\ref{fig.ICO}, are smaller,
rounder and projected further from the star than the bright globulettes \#1 and
\#2. Interestingly, they all fall close to a parabola (drawn as a white curve in
Fig.~\ref{fig.ICO}) with its focus at the star and passing through the IR arc
ahead of HD\,34078. The same is found for the bright globule at the
South-Eastern edge of our field of view. We will further comment on the possible
physical link between CO substructures and the IR arc in
Section~\ref{sec:discussion}.

Figure~\ref{fig.COmoms} displays the spatial distributions of the noise, peak
temperature, line integrated intensities, and centroid velocities of the hybrid
synthesis CO data. The eastern globulette \#1 has a signicantly lower peak
temperature $(\sim1\K)$ than the western one \#2 $(\sim3\K)$. The
eastern/western globulettes have a typical LSR velocity of $\sim5.5$ and
7.4\kms{}, respectively. These velocities approximately correspond to the two
velocities peaks in the IRAM-30m \twCO\Jone{} and \twCO\Jtwo{} spectra obtained
less than $3''$ from the star sightline\footnote{The \twCO\Jtwo{} spectrum is
similar to that presented in \citet{Boisse.2009} but the \twCO\Jone{} is smaller
by a factor $\sim2$. This could come from an erroneous conversion from \Tas{} to
\Tmb{} in the former paper.}, which are shown in the top panel of
Fig.~\ref{fig.spectra}. Our higher resolution dataset thus reveals that the
double-peaked shape of IRAM-30m CO spectra close to the star does not result
from absorption by foreground gas, but from the beam dilution of two almost
distinct globulettes that each contributes to one of the observed CO velocity
peaks. The spatial offset between these two globulettes also readily explains
the East-West velocity gradient observed at $12''$ resolution with the IRAM-30m
by \citet{Boisse.2009}.

The bottom left panel of Fig.~\ref{fig.pv} shows a zoom of the hybrid synthesis
data towards substructures \#1 and \#2. The line of sight towards the star falls
right at the edge of the eastern globulette \#1. In addition, position-velocity
diagrams (\cf{} Fig.~\ref{fig.pv}, upper left and lower right panels) show that
the western substructure \#2 at 7\kms{} has a faint extension to the East, so
that the sightline actually intercepts both velocity components. The
corresponding \twCO\Jone{} emission spectrum towards HD\,34078 at a resolution
of 4\farcs4 is shown in the bottom panel of Fig.~\ref{fig.spectra}.
Table~\ref{tab.LOS} lists the results of a dual Gaussian fit to this emission
spectrum. The two fitted velocity components are in remarkable agreement with
the two narrow absorption components at $\Vlsr \sim 5$ and $\Vlsr \sim 7\kms$
identified in both CH and \CHp{} from high-resolution $(\emr{FWHM}\approx
2\kms)$ optical spectra~\citep[see Fig.~5, 7, and 8 of][]{Rollinde.2003}, even
though the PdBI lobe probes a wider region than the star pencil beam. This
result unambiguously shows that the CO\Jone{} globulettes \#1 and \#2 are
located {\em in front of the star}, and not in its background.

Table~\ref{tab.LOS} also lists the visual extinction to the star that would be
contributed by each CO velocity component assuming a standard \Xco{} conversion
factor $\Wco/N(\Ht) = 2\times10^{20}\pscmpKkms$, and a standard ratio of
$N_\emr{H}/\Av = 1.8 \times 10^{21}\pscm$. The sum is $\Av(\emr{CO}) =
0.3\pm0.1\magn$. This is in excellent agreement with expectations. Indeed, the
reddening suffered by HD\,34078, $\Ebv = 0.52$~\citep{France.2004} implies a
total visual extinction of $\Ebv\times3.1 = 1.6\magn$~\citep[\cf{}][who find a
standard extinction curve towards HD\,34078]{Fitzpatrick.1990}. Out of this,
$1.1-1.5\magn$ comes from atomic HI, whose Ly$\alpha$ and Ly$\beta$ absorption
indicate $N(\emr{HI}) = 2-2.7\times
10^{21}\pscm$~\citep[]{Shull.1985,Boisse.2005}.
This leaves $0.1-0.5\magn$ of extinction by molecular gas on the line of sight.
Therefore, the effective \Xco{} factor in the CO\Jone{} emitting gas appears
close to standard (within a factor 2) and may be used to infer the globulette
masses.

Table~\ref{tab.clumpprops} lists quantitative information about the five small
globulettes. Using standard values for the \Xco{} factor and for
$N_\emr{H}/\Av$, we derived their peak visual extinction from their integrated
emission. They all enter in the category of diffuse clouds, \ie{}, $\Av < 1$
mag. {Following \citet{Solomon.1987}, we yield their ``virial'' masses
as} 
\begin{equation}
	\frac{M_\emr{vir}}{\Msun} = 189 \, \frac{S}{1\pc} \,
	\paren{\frac{\dV}{1\kms}}^2, 
\end{equation}
{with $S$ the globulette
size, and $\dV$ its linewidth. We also computed their luminous masses as}
\begin{equation}
	\frac{M_\emr{co}}{\Msun} = 2\times1.4 \frac{m_\emr{H}}{\Msun}
	\, \Xco \, \Lco 
\end{equation}
{with \Lco{} the globulette luminosity,
$m_\emr{H}$ the proton mass, \Msun{} the solar mass, and \Xco{} the standard
CO-\Ht\ conversion factor, \ie{}, $\sciexp{2}{20}\pscmpKkms$. The numerical
factors account for the \Ht\ mass and the Helium fraction.}
Table~\ref{tab.clumpprops} indicates that the globulettes harbour a very small
mass of molecular gas. Moreover, they are far from being gravitationally bound,
since their luminous mass is at least 3 orders of magnitude smaller than their
``virial'' mass. However, this data alone cannot indicate whether the
globulettes are currently forming, being destroyed, or in an external pressure
confined equilibrium. Their large velocity dispersion $\dV \simeq 1-2.5\kms$
corresponds to an ``effective'' temperature of $\mu\,m_\emr{H}\,\dV^2/k =
300-1500\K$, implying significant non thermal support, or strong internal
velocity gradients (infall, outflow, shear).

Finally, we note that the flux recovered by the PdBI dataset alone inside the
central $50''\times50''$ field of view is only $\sim 1/3$ of the flux measured
with the IRAM-30m telescope. \referee{The recovered fraction even decreases to
$1/4$} when one considers the full field of view imaged at PdBI. This is not a
processing artifact, as we checked that the hybrid synthesis image does recover
the same total flux in the same field of view but at so low brightness level
that it seems hidden in the interferometric noise. Thus, some faint extended
emission must be present in addition to the compact structures measured by the
interferometer~\citep[for a detailed account of a similar effect in another
context, see][]{Pety.2013}. This is consistent with the fact that the deep image
of the \twCO{} \Jtwo{} emission published in~\citet{Boisse.2009} shows faint
emission all over the covered field of view. The low brightness structures
detected here are thus probably part of a more complex extended emission that,
in particular, links the bright compact emission of globulettes \#1, \#2, and
\#3. In other words, these globulettes are probably related (\ie{}, not
distinct) entities.

\section{Discussion} 
\label{sec:discussion}

In this section, we first discuss how the two CO globulettes seen in the
immediate vicinity of the star clarify our understanding of the peculiar
properties of the line of sight towards HD\,34078. We then briefly consider the
possible origin and survival of these substructures. And we finally examine the
implications of our results for the nature of the IR arc and the value of the
wind mass-flux in HD\,34078. For reference, Table~\ref{tab.parameters} lists the
observed and infered parameters.

\subsection{Explaining the peculiar properties of the line of sight towards
HD\,34078} 
\label{ref:discussion:1}

\TabParameters{} In addition to explaining the double-peaked CO emission
profiles and east-west velocity gradient seen in previous IRAM-30m maps (see
previous section), the two small CO substructures found to intercept the line of
sight to HD\,34078 clarify several pending issues regarding the peculiar
properties of this sightline.
\begin{itemize}
	\item {\em The unusually large fraction of highly excited H$_2$
	and the high inferred gas density:} The line of sight to the star passes through
	the {\em edges} of the dense CO substructures \#1 and \#2, where the gas is
	directly exposed to the intense stellar UV flux and forms a hot
	{photo-dissociation region (PDR)}. Such a tangent view will maximize the
	fraction of hot \Ht{} on the line of sight. At the same time, the high density
	in the CO substructures \#1 and \#2 ($\nclump \simeq 10^5\pccm$, see
	Table~\ref{tab.clumpprops}) explains the high $n_\emr{H} \simeq 0.5-5\times
	10^4\pccm$ inferred from PDR models of the excited \Ht{}~\citep[]{Boisse.2005}.
	\item{\em The similar radial velocity and line width among hot and cool \Ht{}
	on the line of sight:} Using a standard \Xco{} factor, the \Wco{} measured
	towards the star inside a 4\arcsec{}-beam (see Table 3) yields an estimated
	total \Ht{} column density of $\simeq 3 \times 10^{20}\pscm$ with an uncertainty
	of a factor 2 (see Section 3). This is consistent with the column density of
	low-excitation \Ht{} at 77\K{} seen in absorption~\citep[$6.4 \times
	10^{20}\pscm$][]{Boisse.2005}. Therefore, the cool \Ht{} on the line of sight
	also appears associated with the CO substructures \#1 and \#2, rather than with
	a foreground translucent cloud unrelated to HD\,34078~\citep [as initially
	proposed in][]{Boisse.2005}. The cool \Ht{} could be located preferentially on
	the ``shadowed'' side of the globulettes facing the observer, while the hot PDR
	would be on the irradiated side facing the star, as proposed by
	\citet{Boisse.2009} in their second scenario.
	\item {\em The high abundance of CH and \CHp{}:} The two components at 5 and
	7\kms{} in the CH and \CHp{} optical absorption profiles can be unambiguously
	associated with CO substructures \#1 and \#2 (see Section 3). The unusually high
	abundance of CH and \CHp{} along this line of sight~\citep[]{Boisse.2009} might
	then be related to the fact that we probe a very peculiar situation. Indeed, the
	globulettes \#1 and \#2 must be located close to the star, at $\simeq
	0.2\pc$~\citep[]{Boisse.2005}, to reproduce the excitation of hot \Ht{} on the
	line of sight. Interaction with the (hot) stellar wind may then be very
	effective in enhancing several species through turbulent mixing or shock
	dissipation, especially \CHp{} whose formation is highly
	endothermic~\citep[]{Boisse.2009}.
	\item {\em The excess reddening towards HD\,34078 compared to the surrounding
	nebula:} Far-UV rocket-borne and FUSE observations of HD\,34078 and its
	neighborhood performed by~\citet{France.2004} revealed that the direct starlight
	is much more reddened than the stellar light scattered by surrounding dust in
	the IC\,405 nebula. This led these authors to propose the existence of an
	obscuring dust globulette in front of the star, with a size smaller than $\sim
	20''$. Our observations provide a direct confirmation of this prediction, since
	substructures \#1 and \#2 are smaller than 10\arcsec. The atomic gas in the hot
	PDR around these globulettes could contribute $\simeq 1\magn$ of extinction on
	the line of sight~\citep[see Table~4 of][]{Boisse.2005}, \ie{}, a sizeable
	fraction of the stellar obscuration.
\end{itemize}
We stress that HD\,34078 was selected for further study because of
its unusually strong absorption features from dense irradiated molecular gas, as
revealed, \eg{}, by FUSE. Therefore,a distance to the star of the close
alignment between the CO substructures \#1 and \#2 and the line of sight seen in
HD\,34078 may well be the result of a selection bias. Indeed, with a transverse
diameter of $4-5''$ and a distance to the star of 0.2\pc{}, these dense
structures subtend an angle of only $5\degr$ as seen from the star, making such
an alignment quite unlikely. This would explain why the line of sight towards
HD\,34078 is so peculiar.

\subsection{Origin and survival of the CO globulettes on the line of sight}
\label{sec:survival}

In order to discuss the possible origin of the small CO globulettes identified
in the vicinity of HD\,34078, it is important to compare their density and
internal pressure to that in the surrounding \ion{H}{ii} region excited by the
star. A rough estimate of the average density in the \ion{H}{ii} region may be
obtained from its radius as traced by bright H$\alpha$ emission, which extends
up to about 27\,arcmin = 4.3\pc{} to the North of the star. Equating to the
Str{\"o}mgren radius (\Rstrom{}) for an O9.5 star, 
\begin{equation}
	\Rstrom
	\simeq 16 \, \paren{L_\star \over 8 \times 10^4 L_\odot} \, \paren{2.5\pccm
	\over \namb}^{2/3} \pc, 
	\label{eq.stromgren} 
\end{equation}
we find $\namb \sim$
{ 20}$\pccm$ (this estimate is only indicative as the \ion{H}{ii} region around
HD\,34078 is quite irregular). Thermal instability in pre-existing density
fluctuations could thus lead to the formation of cooler and denser filaments at
$10-100\K$ and $10^3-10^4\pccm$ (in pressure equilibrium with the \ion{H}{ii}
region). However, this is not sufficient to reach the densities of $\simeq
0.5-1\times10^5\pccm$ estimated for the CO globulettes \#1 and \#2 towards the
line of sight (see Table~\ref{tab.clumpprops}). Hence these globulettes are
overpressured with respect to the surrounding \ion{H}{ii} region.

One attractive hypothesis would be that denser substructures form by compression
in the bow shock driven by the stellar wind into the \ion{H}{ii} region.
A bright IR arc of warm dust is indeed seen in the direction of proper motion of
HD\,34078 (see Fig.~\ref{fig.context}), and attributed to a stellar wind bow
shock ~\citep{France.2007,Peri.2012}. Numerical simulations show that cooling
instabilities will develop behind such bow shocks, creating small globulettes
that are carried along the bow surface~\citep[]{Comeron.1998}. A hint in favor
of this hypothesis is that small globulettes \#3 to \#5 in our map, as well as
the globulette at the south-east of our field of view, all appear to lie close
to the walls of a paraboloid extrapolated from the IR arc (see
Fig.~\ref{fig.ICO} in Section~\ref{sec:observations}). Indeed, we show in
appendix~\ref{sec:projection} that the front side of this paraboloid will
intercept the line of sight at 0.14\pc{} from the star, for an inclination of
the star motion of $-30\degr$ from the plane of the sky. In addition,
\citet{Boisse.2005} showed that the highly excited \Ht{} on the line of sight
(which we find here associated with the edges of globulettes \#1 and \#2) must
be at an actual distance to the star of $\simeq 0.2\pc$. Therefore, globulettes
\#1 and \#2, although projected close to the stellar position, also seem to lie
close to the surface of the paraboloid. However, CO maps over a larger area
upstream and downstream of the IR arc would be needed to confirm this
correlation trend.

An important issue with such a scenario is that UV-derived stellar wind
parameters from \citet{Martins.2005} {suggest} a wind ram pressure at $R
= 0.2$\pc{} from the star, which is several orders of magnitudes too low to
compress the CO globulettes \#1 and \#2 to their high observed density. Namely,
\begin{equation}
	\Pwind/k = \Mrate \Vwind / (4 \pi R^2 k) = 2\,300\Kpccm ,
	\label{eq.pwind} 
\end{equation}
compared to the globulette thermal pressure
\begin{equation}
	\Pther/k = \nclump T \sim 10^6-10^7\Kpccm 
	\label{eq.ptherm}
\end{equation}
for a plausible range of $T = 10-100\K$. Another (related) issue,
pointed out by~\citet{Boisse.2009}, is that the predicted wind bow shock is much
too small compared to the observed size of the IR bow. Possible solutions to
this problem, including a possible underestimate of the wind mass-flux, are
presented and discussed in the next section in the light of our new CO results.

Another process that could contributes to the compression and
pressure-confinement of the globulettes is radiation-driven implosion. This
eventually leads to the formation of a ``cometary globule'' with a
photoevaporating dense head and a tail pointing away from the star~\citep[see,
\eg{},][]{Lefloch.1994}. The direction of elongation of substructures \#1 and
\#2 is reminiscent of such a geometry. Taking into account their convolution by
the beam size, their small observed transverse radius $\sim 2'' = 0.005\pc$ is
close to the predictions of Eq.~45 of~\citet{Lefloch.1994} for a cometary
globule with the observed mass $\sim 5 \times 10^{-3}\Msun$ and linewidth $\dV
\simeq 2\kms$ (corresponding to a non-thermal support with an ``effective''
temperature $\sim 1000\K$), located at 0.2\pc{} from an O9.5 star of total
ionizing flux $\Sstar =$ { 4.7} $\times 10^{47}$ s$^{-1}$ (for \Lstar = { 8}
$\times10^4$ \Lsun). Combining Eqs.~45 and~50 of \citet{Lefloch.1994}, the
lifetime of a cometary globule $\timecg$ until complete photoevaporation may be
simply expressed as a function of its radius and linewidth as 
\begin{equation}
	\timecg = 2.4 \times 10^4\yr \, \paren{\Rcg \over 0.005\pc} \sqrt{2\kms \over
	\dV}.
\end{equation}
By that time, the star will have travelled 3.6\pc{} (or
$0.4\deg$) at a velocity of 150\kms. Therefore, globulettes \#1 and \#2 seem
able to survive against the harsh UV flux of the star if they are in the
cometary globule phase.

\subsection{Effect of radiation pressure and the weak wind problem}
\label{sec:bowshock}

We now examine the implications of our results on the nature of the bright IR
arc observed in the direction of propagation of HD\,34078, and on the value of
the wind mass-flux in this star.

\subsubsection{Stationary bow shock}

The IR arc around HD\,34078 has been traditionally attributed to a bow shock
created by the stellar wind \citep{France.2007,Peri.2012}. However,
\citet{Boisse.2009} recently questioned this interpretation, noting that the
observed star-apex distance \Robs $\simeq 15\arcsec = 1.2\times 10^{17}\cm =
0.04\pc$ was 100 times too large\footnote{The star-apex distance of $1.7 \times
10^{15}\cm$ used in \citet{France.2007} is erroneous by two orders of
magnitude.} compared to the expected wind standoff distance, $\Rwind$
\begin{equation}
	= \sqrt{\Mrate \Vwind \over 4 \pi \rho_a \Vstar^2} \simeq
	3.5\times 10^{-4}\pc \sqrt{500\pccm \over \namb} \paren{100\kms \over \Vstar}
	\label{eq.Rwind} 
\end{equation}
if one adopts the mass-loss rate and wind
velocity inferred from UV line modeling, $\Mrate = 10^{-9.5}\Msunpyr$ and
$\Vwind = 800\kms$~\citep{Martins.2005}, a stellar motion of $\Vstar = 100\kms$,
and the ambient H nucleus number density $\namb = \rhoamb / (1.4 m_{\rm H}) \sim
500\pccm$ estimated from CI and C$_2$ absorption
lines~\citep{Boisse.2005,Federman.1994}.

Our high-resolution mapping results partly alleviate the problem noted by
\citet{Boisse.2009} since they reveal that the line of sight to the star is
crossing the envelope of small dense CO globulettes, so that the density of
$\sim 500\pccm$ probed in CI and C$_2$ may not be representative of the
large-scale ambient gas. Indeed, the size of the \ion{H}{ii} region rather
suggests a mean $\namb \sim$ { 20\pccm} (see Section~\ref{sec:survival}),
consistent with the lack of bright extended CO emission upstream of the IR arc
in the IRAM-30m data. In addition, correcting the observed star-apex distance,
\Robs, for projection effects gives a slightly smaller true value $\Rapex =
0.035\pc$ (see Appendix~\ref{sec:projection}). At the same time, the stellar
peculiar velocity has been recently constrained to $\Vstar \simeq
150_{-50}^{+30}\kms$~\citep{Tetzlaff.2011}. Thus, even with this lower ambient
density, the predicted wind standoff radius $\Rwind =$ { 1--2} $\times
10^{-3}\pc$ is still $\sim 20$ times smaller than the true star-apex distance
\Rapex.

The same issue was recently encountered in the IR bow shock driven by the
runaway star $\zeta$\,Oph~\citep{Gvaramadze.2012}, which has the same luminosity
and spectral type as HD\,34078. To solve this problem in the case of
$\zeta$\,Oph, \citet{Gvaramadze.2012} proposed that the actual wind mass-loss
rate in $\zeta$\,Oph is much larger than inferred from UV lines.
Indeed, there is still ongoing debate on the true wind mass-fluxes in late O
stars, as UV line analyses yield typically 100 times smaller values than
previous H$\alpha$ line fitting and theoretical prescriptions --- the so-called
``weak-wind problem''~\citep[see, \eg{},][for a review and possible
explanations]{Mokiem.2007}. In the case of HD\,34078, we would obtain a
star-apex distance $\Rapex = 0.035\pc$ with $\Mrate = { 1-3}
\times10^{-7}\Msunpyr$ for $\namb \sim$ { 20} $\pccm$ and $\Vstar =
100-180\kms$. This is { 300-1000} times larger than the UV-determined \Mrate{}
value, but only a few times larger than the \Mrate{} inferred in
$\zeta$\,Oph~\citep{Gvaramadze.2012}, and in good agreement with the theoretical
value of { 2} $\times 10^{-7}\Msunpyr$ predicted by the prescription\footnote{We
could not reproduce the smaller theoretical \Mrate\ quoted in
\citet{Martins.2005} for this prescription.} of \citet{Vink.2001} using the
stellar parameters in \citet{Martins.2005}.
The latter theoretical value was in fact used by \citet{Peri.2012} to estimate
the ambient density ahead of HD\,34078 from the IR bow shock size (for which
they adopted a slightly larger apex distance than us, of 0.06\pc{}). With this
increased \Mrate{}, the wind ram pressure at 0.2\pc{} would become $\simeq
10^6\Kpccm$, comparable to the CO globulette thermal pressure, and may
contribute to their compression at the interface. Yet, a puzzling fact in this
picture is the low globulette radial velocity, which is unexpected after
compression by a fast bow shock at 150\kms.

\citet{Boisse.2009} mentioned two other effects that might help to increase the
bow standoff distance without increasing \Mrate{}, namely: (1) a non-stationary
bow shock, (2) radiation pressure on grains. Their estimate of the ratio of
radiative to wind ram pressure in HD\,34078, $\Prad/\Pwind$ $\sim 10^{-3}$, was
incorrect and lead them to erroneously conclude that radiation pressure was
negligible, favoring option (1). The correct value $\Prad/\Pwind$ { =
\Lstar/(c\Mrate \Vwind)} $\sim$ { 6600}, with 
\begin{equation}
	\Prad/k =
	\frac{\Lstar}{4\pi c R^2 k} = \sciexp{1.5}{7} \K\pccm 
	\label{eq.prad}
\end{equation}
suggests that radiation pressure may actually have a significant
effect. Given the importance of determining accurate wind mass-fluxes in late O
stars such as HD\,34078, we thus re-examine and quantify both options in order
to see whether they may explain the large observed star-apex distance, as well
as the apparent good correlation of the IR arc with the CO globulettes, while
conserving the UV-determined \Mrate{} value.

\subsubsection{Non-stationnary bow shock}

In this scenario, we would be witnessing a transient situation where the bow
shock was initially propagating into a very low density medium at $\namb \leq
0.02\pccm$ (so that $\Rwind \ge 0.04\pc$), and is just now entering a much
denser region to the North at 10\pccm{}. The wind ram pressure at the interface
would now be much smaller than the ambient pressure, so that the apex is
essentially stalled at its current position. It will remain so until the
star-apex distance has decreased to a few times its new equilibrium standoff
value $10^{-3}\pc$. The needed time is thus approximately the time for the star
to propagate over $0.04\pc$, \ie{}, 260~yr at $150\kms$. This is so short that
it seems very unlikely. It appears even more implausible to believe that
$\zeta$\,Oph could also be caught in the same transitory situation. 

\subsubsection{Radiation pressure on dust grains without gas drag} 

\citet{Artymowicz.1997} have described the trajectory of dust grains {\em
without gas drag} near a bright moving star and showed that they are repelled
outside of an avoidance region with a paraboloidal shape focussed at the star.
The standoff avoidance radius of the parabola for a grain of radius $s$ is given
by 
\begin{equation}
	\Rrad(s) = 2 \bracket{\beta(s)-1}\, \frac{G
	{\Mstar}}{\Vstar^2}, 
\end{equation}
where $\beta(s) \propto \Lstar/\Mstar$ is
the ratio of radiation pressure to gravity for that grain \citep[see Eq.~4
in][]{Artymowicz.1997}. For an O star, $\beta(s) \gg 1$ (negligible gravity) and
the dependences on \Mstar{} in $\beta$ and \Rrad{} cancel out, so that $\Rrad(s)
=$ 
\begin{equation}
	0.24\pc \paren{\Lstar \over 8 \times 10^4\Lsun}
	\paren{150\kms \over \Vstar}^{2} \paren{Q_{\rm pr}(s)/s \over 30 \mum^{-1}}
	\paren{2.2\gpccm \over \rhobulk}.
	\label{eq.avoidance} 
\end{equation}
{ The adopted value of \rhobulk{} is for a
50\%-50\% compact mixture of astronomical silicates and graphite
\citep{Artymowicz.1997}. The Planck-averaged value of $Q_{\rm pr}(s)/s$ for an
incident black-body at $T_{\star} = 31,000\K$ (appropriate for an O9.5 star)
varies strongly with grain radius, from $\simeq 85$ at small radii $s \le
0.01\mum$ to $\simeq 6$ at $s = 0.3\mum$~\citep{Laor.1993}.} Hence, the dust
avoidance radius $\Rrad(s)$ also varies strongly over this size range, from {
0.7 to 0.05 pc}. Modeled IR images taking into account the variation of
$\beta(s)$ over grain size for an MRN distribution are presented by
\citet{Gaspar.2008} and show a broadening of the IR arc that seems consistent
with the 24\mum{} Spitzer image in HD\,34078.
The value of $Q_{\rm pr}(s)/s = 30$ in the above scaling is a mass-weighted
average over a standard MRN size distribution with $n(s) \propto s^{-3.5}$ in
the range $[0.005\mum, 0.25\mum]$, and corresponds to a grain of size
$\overline{s}=0.05\mum$. The corresponding mass-weighted avoidance radius,
$\overline{\Rrad}$ { = 0.24 pc} is a factor { 7} larger than the star-apex
distance of the IR arc in HD\,34078. Given the { relative} uncertainty in
\Vstar{} (a factor 1.5) and in dust properties (composition, { size
distribution}, volume density and porosity), this might still be considered a
promising agreement\footnote{ As the motion of H34078 is mainly in the plane of
the sky, \Vstar{} is to first order $\propto D$. Furthermore, \Lstar\ is
$\propto D^2$ for a given apparent magnitude, reddening, and spectral type.
Therefore $\Rrad(s) \propto \Lstar/\Vstar^2$ is independent of the assumed
distance.}.

A strong caveat of this model is that it cannot explain the distortion of the IR
arc observed in the 24\mum{} image from a perfect paraboidal shape pointing in
the direction of motion of the star (Fig.~\ref{fig.context} shows that the
eastern and western wings of the arc are better reproduced with a different
\Robs, of 20\arcsec and 10\arcsec, respectively). Indeed, unlike the wind
standoff radius \Rwind{}, the dust avoidance radius \Rrad{} { in
Eq.~\ref{eq.avoidance}} is {\em independent of the ambient density} so that a
parabola aligned with \Vstar{} should result, whatever the inhomogeneities
present in the medium ahead of the star. Another issue is that the IR bow shock
distance in runaway stars of similar spectral types and dust properties should
vary as $\Rrad \propto \Lstar / \Vstar^2$ regardless of ambient density. This
relation predicts that the IR bow shock in the O9.5 star $\zeta$\,Oph, where
$\Lstar \simeq 6\times 10^4\Lsun$ and $\Vstar \simeq
25\kms$~\citep{Gvaramadze.2012}, should be { 12--40} times larger than in
HD\,34078 (where { \Lstar = 8 $\times 10^4\Lsun$ and} $\Vstar = 100-180\kms$),
whereas one observes only a factor 4.

The above determination of \Rrad{} assumes negligible gas drag, \ie{}, that
grains and gas are totally decoupled. Therefore, the gas will continue to
approach the star until it meets the wind bow shock defined by the (tiny) value
of $\Rwind \simeq 10^{-3}\pc$ for the UV-determined \Mrate{}. The fact that the
CO globulettes in our maps appear to lie on the same surface as the dust
avoidance parabola defined by the IR arc might then be understood if they
quickly photodissociate when entering the avoidance region, as they leave the
dust behind. However, the density in the CO globulettes appears too high for
dust and gas to remain decoupled. The distance $d(s)$ over which a grain of
radius $s$ and bulk density \rhodust{} will sweep up its own mass in gas is
\begin{equation}
	d(s) = 0.04\pc \paren{100\pccm \over \namb} \paren{s \over
	0.1\mum} \paren{\rhodust \over 2.23\gpccm}.
\end{equation}
We thus expect significant dust-grain coupling to occur on
observed scales if the density exceeds a few $100\pccm$, which is certainly the
case inside the dense CO globulettes even for the largest grains.

\subsubsection{Radiation pressure on dust grains with gas drag (optically thin
case)} 

We now consider how the avoidance radius created by radiation pressure on grains
would change in the presence of gas drag. In the limiting case of perfect
gas-grain coupling, the radiation pressure force on grains (assumed optically
thin to the stellar radiation) will be entirely transferred to the gas. As
demonstrated in Appendix~\ref{sec:dust-drag}, the dynamics of the coupled fluid
upstream of the avoidance radius remains similar to that for pure dust grains,
except that the effective ratio of pressure to gravity $\overline{\beta}$
(averaged over the grain size distribution) is divided by the gas-to-dust ratio
$(\rhogas/\rhodust) \simeq 100$ because of the gas inertia. The resulting
standoff radius is given by the mass-weighted dust avoidance radius
$\overline{\Rrad}$ reduced by the same factor of 100, \ie{}, 
\begin{equation}
	\Rthin \simeq {\mathbf 2.4} \times 10^{-3}\pc 
	\label{eq.rthin} 
\end{equation}
for the parameters of HD\,34078. Unlike pure dust grains, the dusty gas will not
``bounce'' off the avoidance parabola but undergo a shock front and slide along
the parabola walls. The additional wind ram pressure acting on the fluid will
increase this shock standoff distance to a value close to the maximum between
\Rthin{} and \Rwind{} (see Appendix~\ref{sec:dust-drag}).

The relative importance of radiation pressure versus wind ram pressure in
defining the standoff radius in this (optically thin) case is quantified by the
ratio $\Rthin/\Rwind$, rather than by the ratio $\Prad/\Pwind$. Since \Rthin{}
is even smaller than \Rwind{} for the standard value of \Mrate{}, radiation
pressure on grains in the dense CO globulettes (assumed optically thin) does not
explain, alone, their observed distribution along the IR arc, which has a { 15}
times larger standoff distance. \footnote{ In a paper accepted while this work
was under revision, \citet{Ochsendorf.2014} investigate the case of imperfect
gas-drag and argue that it could explain the IR arc sizes in runaway stars such
as $\sigma$\,Ori and $\zeta$\,Oph if grains in \ion{H}{ii} regions are much less
charged than usually assumed. Like us, however, they still predit too small a
size for the infrared arc in HD 34078 where \Vstar\ is large and gas-grain drag
is dominated by direct collisions rather than by coulomb interactions.}

\subsubsection{Radiation pressure on dust grains with gas drag (optically thick
case)} 

The situation may change if the dense globulettes with good gas-grain coupling
become optically thick to the stellar UV radiation. We note that the maximum
effect will be obtained by replacing \Pwind{} by \Prad{} in { the expression of
\Rwind{}, \ie{}, in Eq.~\ref{eq.Rwind}}. Such an upper limit would be reached if
the incoming globulettes do not feel the radiation pressure of the star until
they get to the bow shock, because of the screening by other optically thick
globulettes closer to the bow surface. For the parameters of HD\,34078 and
$\Vstar = 150\kms$, this would give 
\begin{equation}
	\Rthick = \Rwind \,
	\sqrt{(\Prad/\Pwind)} \simeq 0.1\pc.
	\label{eq.rthick} 
\end{equation}

The { similarity with the value of} $\overline{\Rrad}$ is purely coincidental,
since the dependence on stellar parameters and ambient density is different,
with $\Rthick \propto \sqrt{\Lstar/\namb} \times (1/\Vstar)$. Such a scaling
would better explain the ratio of 4 between the IR apex size in $\zeta$\,Oph and
HD\,34078, if the ambient densities are similar. And the dependence on \namb{}
could explain bow shock distortions with a density (or magnetic pressure)
gradient inclined with respect to the star proper motion, in contrast to the
case of pure dust avoidance.
However, the assumption of an optically thick bow shock is clearly too extreme
as the star is able to ionise a large \ion{H}{ii} region ahead of it. A more
realistic modeling is outside the scope of the present paper, but would be
useful.

\section{Conclusions} 
\label{sec:conclusions}

We described the calibration and construction of the \twCO{} \Jone{} imaging at
$\sim10^{-3}\pc$ ($\sim4.4''$) of the $\sim 0.31 \times 0.26\pc$ $(\sim
120''\times100'')$ toward the runaway O star HD\,34078, using observations from
both the PdBI and IRAM-30m telescopes. The IRAM-30m data mainly features two
unresolved globules: One at the south eastern edge of the observed field of
view, not discussed here, and a second one around the star sightline. At the
PdBI resolution, the latter appears to be composed mainly of 2 bright (1 and
3\K{} peak temperature) and compact (size $\le 10''$ or $0.026\pc$) globulettes
linked together by an extended faint emission that amounts to two third of the
total flux.

The star sightline clearly intercepts the edge of globulette \#2.
However, the spectrum in the direction of the star is double-peaked indicating
that the star sightline also intercepts globulette \#1. These globulettes are
responsible for the absorption lines from CH, \CHp{} and cold \Ht{} at 77\K{},
and also explain the large amounts of dense excited \Ht{} at 350\K{} on this
sightline, the latter probably tracing a PDR on the illuminated face of the
globulettes. The measured CO column density in this direction is compatible with
the stellar reddening measured in far-UV and visible. The globulettes are small
enough for the surrounding reflexion nebula to be less reddened than the star
\citet{France.2007}. The imaged CO globulettes appear to approximately lie along
the parabola walls. They are clearly not gravitationally bound. They may be
pressure confined and they probably result from the interaction between the star
and the preexisting diffuse gas (\eg{}, thermal instabilities in the bow shock).

We quantified the actions of two competing processes in the interaction between
the star and the preexisting diffuse nebula, IC\,405. The first one is
ram-pressure due to the high velocity star wind. The second one is the radiative
pressure (optically thin case) on the dust grains that entrain the gas through
friction.

Neither a non-stationary bow shock nor the effect of radiation pressure (in the
optically thin limit) can explain at the same time (1) the observed large size
of the IR arc in HD\,34078, (2) its distortion from a perfect parabolic shape,
(3) its size ratio of 4 compared to that around $\zeta$\,Oph, and (4) its
spatial correlation with dense CO globulettes.
The most straightforward explanation for these 4 properties appears to be that
the wind mass-flux is { 300-1000} times larger than indicated by UV lines, and
close to the theoretical prescription of~\citet{Vink.2001}. The effect of
radiation pressure on optically thick globulettes would be important to
investigate but may also prove insufficient. Indeed, the low radial velocities
observed in the CO globulettes appear puzzling for a steady-state wind bow
shock.

A study over a wider field of view in both CO and \Halpha{}, including regions
both upstream and downstream from the IR arc, would be important to obtain
additional constraints on the formation process and dynamical state of the CO
globulettes.
\begin{acknowledgements}
	This work has been funded by the grant
	ANR-09-BLAN-0231-01 from the French {\it Agence Nationale de la Recherche} as
	part of the SCHISM project (http://schism.ens.fr/). PG acknowledges support from
	the ERC Starting Grant (3DICE, grant agreement 336474) at the end of this work.
	JP and PG thanks H.S.~Liszt, R.~Lucas, and A.~Witt for their encouragements
	during this work.
\end{acknowledgements}

\bibliographystyle{aa} 
\bibliography{hd34078-pdbi} 

\appendix{}

\section{Projection effects} 
\label{sec:projection}

\FigProjection{}

Projections effects must be taken into account when interpreting the observed
contours of the paraboloid, because the star velocity vector $\vec{\Vstar}$ lies
out of the plane of the sky. Let $\incli$ be the inclination of the velocity
vector with respect to the plane of the sky, with $\incli<0$ when the star is
receding. In our case, $\incli \approx - 30\degr$. Figure~\ref{fig.projection}.a
shows the natural reference frame, named O$xyz$, associated with the star
motion: O coincides with the star position, O$y$ is parallel to $\vec{V_*}$, and
O$x$ is in the plane of the sky. Adopting the star-apex distance OA as unit
length, the equation of the paraboloid in this frame is $y = 1 - (x^2+z^2)/4$.

The observer frame, named OXYZ, is rotated by $-\incli$ about O$x$=OX so that OZ
is pointing away from the observer, and OY is the projection of $\vec{V_*}$ in
the plane of the sky. For $\incli \not= 0$, the equation of the paraboloid in
the OXYZ frame can be obtained by performing the appropriate coordinate
rotation. This yields 
\begin{equation}
	-Z\sin \incli + Y\cos \incli = 1 - \frac
	{(Z\cos\incli + Y\sin\incli)^2 +X^2}{4}.
	\label{eq.paraboloid} 
\end{equation}

This allows us to define the coordinates of three specific points of the
parabola. The apparent apex A$'$ is the point in the $X=0$ plane where
${dY}/{dZ}=0$. We obtain $Z= \sin \incli/\cos^2\incli$ and the apparent
star-apex distance $Y_{\rm max}= 1/\cos \incli$. The two points of the parabola
intersecting the line of sight are found by setting $Y=0$ and $X=0$ in
Eq.~\ref{eq.paraboloid}. The two solutions are 
\begin{equation}
	Z_{\pm} =
	\frac{2 \sin\paren{\incli \pm 2}}{\cos^2\incli} = \frac{2 \sin\paren{\incli \pm
	2}}{\cos\incli} Y_{\rm max}.
\end{equation}

Figure.~\ref{fig.projection}.b shows the shape of the paraboloid as seen in
projection in the plane of the sky. It is obtained by setting $dX/dZ = 0$ in
Eq.~\ref{eq.paraboloid}. This yields 
\begin{equation}
	Y\cos \incli = 1 - \frac
	{X^2 \cos ^2\incli }{4}.
\end{equation}
This still is a parabola focused in O. This curve intersects the
X-axis at a projected distance from the star $X=\pm \,2/\cos \incli = \pm 2
Y_{max}$.
The projected parabola thus has an aspect ratio identical to the original one.

For $\incli = -30^{\circ}$ and an observed star-apex distance of OA$'' =
0.04\pc$, we infer the true star-apex distance OA = OA$''\,\cos\incli =
0.035\pc$. The distance from the star to the intersection point C closest to the
observer is then OC = 2OA$'' \,(\sin\incli - 1)/\cos\incli = 0.14\pc$.

\section{Radiation pressure on grains with gas-drag} 
\label{sec:dust-drag}

\newcommand{\mybeta}{\emm{\overline\beta}} 
\newcommand{\myalpha}{\emm{\alpha}}

The dust-gas collisions exchange momentum between the gas and dust components
and thus transfer the radiative impulsion felt by the grains to the gas. The
dust momentum equation reads 
\begin{equation}
	{\rm D}_{t}
	\paren{\rhodust\boldsymbol{V}_\emr{d}} =
	(\mybeta-1)\,\frac{G\Mstar}{R^{3}}\boldsymbol{\hat{R}}\,\rhodust + \Fdrag,
\end{equation}
where ${\rm D}_{t}$ denotes the Lagragian derivative $ 
\partial_t
+ V \dot \nabla$, $\boldsymbol{V}_\emr{d}$ and $\rhodust$ are the dust velocity
and density per unit volume of gas, $R$ is the distance from the star with its
unit vector $\boldsymbol{\hat{R}}$, \mybeta{} is the ratio between the radiative
force on the grains and the gravitational force (\cf{} Artymowicz \& Clampin
1997) averaged over the grain size distribution, and \Fdrag{} is the momentum
transfer between gas and dust due to dust-gas friction. The corresponding
equation for the gas fluid is 
\begin{equation}
	{\rm
	D}_{t}\paren{\rhogas\boldsymbol{V}_\emr{g}} + \boldsymbol{\nabla}p =
	-\frac{G\Mstar}{R^{3}}\boldsymbol{\hat{R}}\,\rhogas - \Fdrag, 
\end{equation}
with $\rhogas$ and $\boldsymbol{V}_\emr{g}$ the gas mass density and velocity,
and $p$ the thermal pressure. If we sum up these two equations to get the
evolution for the total momentum, we get 
\begin{equation}
	{\rm
	D}_{t}\paren{\rhogas\boldsymbol{V}_\emr{g}} + {\rm
	D}_{t}\paren{\rhodust\boldsymbol{V}_\emr{d}} + \boldsymbol{\nabla}p =
	(\mybeta\rhodust-\rhogas-\rhodust)\,\frac{GM}{R^{3}}\boldsymbol{\hat{R}}.
\end{equation}

We now assume 1) that the gas ram pressure dominates both its thermal pressure
$p$ and the ram pressure of the dust; 2) that the system has reached a
steady-state with $ 
\partial_t\equiv 0$; 3) that we are on the star-apex axis,
an axis of symmetry for the system; { 4)} that $ 
\partial_x
(\boldsymbol{V}_\emr{g})_{x} \simeq 0$ where $x$ is any direction orthogonal to
the star-apex axis (\ie{}, the slight divergence of the incoming flow is
negligible at the apex). The total momentum equation then yields
\begin{equation}
	\partial_{R}\paren{\rhogas \Vgas^{2}} =
	\myalpha\frac{G\Mstar}{R^{2}}\rhogas, 
	\label{eq.mom} 
\end{equation}
where
$\myalpha=\mybeta\rhodust/\rhogas-1$. Similarly, the continuity equation yields
\begin{equation}
	\partial_{R} \paren{\rhogas \Vgas}=0.
	\label{eq.mass} 
\end{equation}
Under these approximations, the gas behaves
exactly like dust with an effective \mybeta{} lowered by the factor
$\rhodust/\rhogas$. We use \rhoamb{} and \Vstar{} as the corresponding values of
\rhogas{} and \Vgas\ far from the star. The above conservation equations
(\ref{eq.mom}) and (\ref{eq.mass}) become 
\begin{equation}
	\rhogas \Vgas =
	\rhoamb \Vstar 
	\label{eq.cmass} 
\end{equation}
\begin{equation}
	\mbox{and} \quad
	\frac{1}{2} \Vgas^{2} + \myalpha\frac{G\Mstar}{R} = \frac{1}{2} \Vstar^{2}
	\label{eq.cmom} 
\end{equation}
which constrain completely the profile of the gas
mass density and velocity in the upstream gas under the repulsive effect of
radiation pressure. We define $\Rthin = {2\myalpha G\Mstar}/\Vstar^{2}$ the
corresponding ``hybrid'' gas+grain avoidance radius where $\Vgas = \Vdust = 0$
in the presence of grain-gas coupling.

We then look for the corresponding steady-state stand-off radius $\Rapex$ of the
bow shock created by the stellar wind. Its position is determined by the balance
(in the reference frame of the star) between the ram pressures in the wind and
in the incoming gas stream 
\begin{equation}
	\rhowind\Vwind^{2}=\rhogas
	\Vgas^{2}.
	\label{eq.balance} 
\end{equation}
We denote as \Rwind{} the wind stand-off
radius in the absence of dust drag, \ie{}, when $\Vgas = \Vstar$. The radiation
pressure on dust lowers the velocity of the upstream gas to $\Vgas<\Vstar$,
hence the pressure balance will be obtained for $\Rapex >\Rwind$. Similarly, if
$\Vwind > 0$, the ram pressure balance will occur at $\Vgas > 0$, hence $\Rapex
>\Rthin$.
Thus, \Rapex{} will be larger than both \Rwind{} and \Rthin{}. Combining
Eqs.~\ref{eq.cmass}, \ref{eq.cmom}, and~\ref{eq.balance}, we obtain
\begin{equation}
	\paren{\frac{\Rwind}{\Rapex}}^2=\sqrt{1-\frac{\Rthin}{\Rapex}}.
\end{equation}
Finally, with $x=\Rwind/\Rapex$, this last equation becomes
\begin{equation}
	x^{4} + \frac{\Rthin}{\Rwind}x - 1 = 0, 
\end{equation}
which
completely determines \Rapex{} from \Rthin{} and \Rwind{}.

The stand-off radius is only slightly larger than the largest value between
\Rthin{} and \Rwind{}. When $\Rwind \gg \Rthin$, the gas feels almost no dust
drag due to radiation pressure before entering the shock and $\Rapex \simeq
\Rwind$. Conversely, when $\Rthin \gg \Rwind$, the wind ram pressure is quickly
negligible ahead of the avoidance radius and $\Rapex \simeq \Rthin$. When
$\Rwind = \Rthin$, we have $\Rapex \simeq 1.5\Rwind$.

\end{document}